\documentclass[a4paper,11pt]{article}
\pdfoutput=1 

\usepackage{jheppub} 

\usepackage{multirow, graphicx,amssymb,url,mathrsfs,amsmath}
\usepackage{subfig}
\usepackage{amsxtra,amstext,latexsym,dsfont,amsfonts,amssymb}
\usepackage{color,eucal}
\usepackage[dvipsnames]{xcolor}
\usepackage{float}
\usepackage{slashed,comment}
\usepackage{cases}
\usepackage{xspace}

\usepackage{mathtools}
\usepackage{tikz} 
\usepackage{tzplot}
\usepackage[compat=1.1.0]{tikz-feynman}
\usepackage{feynmf}

\usepackage{hyperref}
\hypersetup{colorlinks=true,allcolors=blue}
\usetikzlibrary{decorations.pathmorphing}

\tikzset{snake it/.style={decorate, decoration=snake}}

\newcommand{\e}{\epsilon}      \newcommand{\D}{\Delta}
      
\renewcommand{\d}{\delta}      
\renewcommand{\l}{\lambda}     \renewcommand{\L}{\Lambda}
      \renewcommand{\t}{\tau}

\newcommand{\g}{\gamma}      
 
\newcommand{\m}{\mu}         
\newcommand{\n}{\nu}

\newcommand{\cala}{\mbox{${\cal A}$}} \newcommand{\calb}{\mbox{${\cal B}$}}
 
 \newcommand{\calf}{\mbox{${\cal F}$}}

 \newcommand{\call}{\mbox{${\cal L}$}}
\newcommand{\calm}{\mbox{${\cal M}$}} 
 \newcommand{\calp}{\mbox{${\cal P}$}}

 \newcommand{\calv}{\mbox{${\cal V}$}}

\newcommand{\nn}{\nonumber}

\newcommand{\ra}{\rightarrow}

\newcommand{\Poincare}{Poincar\'e\xspace}

\newcommand{\lb}{\left[}
\newcommand{\rb}{\right]}

\title{Wedge Holographic Complexity in Karch-Randall Braneworld}

\author[a]{Yichao Fu}
\author[a,b]{and Keun-Young Kim}
\emailAdd{yichao.fu@gm.gist.ac.kr}
\emailAdd{fortoe@gist.ac.kr}

\affiliation[a]{Department of Physics and Photon Science, Gwangju Institute of Science and Technology, \\
123 Cheomdan-gwagiro, Gwangju 61005, Korea}

 \affiliation[b]{Research Center for Photon Science Technology, Gwangju Institute of Science and Technology, 123 Cheomdan-gwagiro, Gwangju 61005, Korea}

\abstract{
We investigate holographic complexities in the context of wedge holography, focusing specifically on black string geometry in AdS$_3$. The wedge spacetime is bounded by two end-of-the-world (EOW) branes with intrinsic Dvali-Gabadadze-Porrati (DGP) gravity. In line with this codimension-two holography, there are three equivalent perspectives: bulk perspective, brane perspective, and boundary perspective. Using both the ``Complexity=Volume'' (CV) and ``Complexity=Action'' (CA) proposals, we analyze the complexity in wedge black string geometry in the tensionless limit. By treating the branes as rigid, we find the late-time growth rates of CV and CA match exactly across bulk and brane perspectives. These results are consistent with those from JT gravity, with additional contributions from the intrinsic gravity of the branes. For fluctuating branes, we find that the late-time growth rates of CV and CA match between bulk and brane perspectives at the linear order of fluctuation. The CV results exhibit $\frac{\phi_h^2}{\phi_0}$ corrections from fluctuations, consistent with the findings in previous work. Moreover, the CA results reveal an additional constant term in the fluctuating branes case. We provide an interpretation of this in terms of gravitational edge mode effects. The distinct corrections arising from fluctuations in the CA and CV proposals suggest that the CV proposal is more sensitive to geometric details. Furthermore, we discuss these results in relation to Lloyd's bound on complexity, their general time dependence, and the effects of fluctuations. 
}

\begin{document}
\maketitle

%
\section{Introduction}
AdS/CFT correspondence~\cite{Maldacena:1997re, Gubser:1998bc, Witten:1998qj} offers a powerful framework for understanding quantum gravity, one of the most important challenges in modern theoretical physics. It connects observables in $d$-dimensional bulk spacetime to those in codimension-one boundary field theory. Later developments proposed a holographic duality between bulk gravity and boundary CFTs, now known as AdS$_d$/BCFT$_{d-1}$~\cite{Takayanagi:2011zk, Fujita:2011fp, Karch:2000gx}. This form of holography is naturally realized in the Karch-Randall braneworld~\cite{Randall:1999ee, Randall:1999vf, Gubser:1999vj, Karch:2000ct}, where an end-of-the-world (EOW) brane is present in the bulk spacetime described by (semi)classical gravity, see \cite{Geng:2020fxl, Geng:2020qvw, Geng:2021hlu, Geng:2021iyq, Geng:2022dua, Geng:2021mic, Geng:2021wcq, Geng:2020kxh, Geng:2023zhq, Yadav:2023qfg, Wen:2024uwr, Lin:2023ajt, Aguilar-Gutierrez:2023tic, Aguilar-Gutierrez:2023ccv, Aguilar-Gutierrez:2023zoi, Kim:2023adq, Chang:2023gkt, Yadav:2023sdg, Li:2023fly, Kanda:2023zse, Miao:2023unv, Hu:2022zgy, Geng:2024xpj} for recent work on entanglement and complexity in this background. Furthermore, an additional EOW brane can be introduced in the bulk, intersecting with the first at the asymptotic boundary, forming a wedge near the boundary. This codimension-two holography is referred to as wedge holography~\cite{Akal:2020wfl, Geng:2020qvw, Bousso:2020kmy, Miao:2020oey}, which posits a holographic duality between a wedge gravitational theory and a codimension-two CFT located at the corner of the wedge. Later, It was carefully formulated in \cite{Geng:2020fxl, Geng:2021iyq}.  Recent work has demonstrated that in the AdS$_3$ wedge spacetime, with two fluctuating branes, the gravitational action reduced to JT gravity in the brane tensionless limit~\cite{Geng:2022tfc, Geng:2022slq}.

On the other hand, studying the black hole interior quantum mechanically is always crucial and compelling, particularly in the context of the black hole information paradox. While the  AMPS~\cite{Almheiri:2012rt} proposal introduced the idea of a firewall, an alternative resolution was offered by Maldacena and Susskind through the ER=EPR conjecture~\cite{Maldacena:2013xja}, which states that two entangled states are connected by an Einstein-Rosen~(ER) bridge. To analyze the entanglement of these states, one can compute the entanglement entropy holographically using the Ryu-Takayanagi~(RT) formula~\cite{Ryu:2006bv}. However, it was later suggested that entanglement entropy alone is insufficient to describe the late-time growth of the ER bridge after the rapid thermalization of the black hole. This observation has led to the exploration of the computational complexity of states, or its dual, holographic complexity \cite{Susskind:2014moa}.

In the spirit of bulk/boundary correspondence, the computational complexity of states in the boundary field theory is dual to the holographic complexity in bulk gravity. In the context of two-sided black holes, the holographic complexity is proposed to be equivalent to the length or volume of an extremal codimension-one hypersurface connecting the boundary states, known as the ``Complexity=Volume'' (CV) proposal~\cite{Susskind:2014rva, Stanford:2014jda}. Later, an alternative proposal-known as the ``Complexity=Action'' (CA) proposal~\cite{Brown:2015bva}-suggested that the complexity of boundary states is dual to the on-shell gravitational action within a bounded ``Wheeler-DeWitt''~(WDW) patch in the bulk. Due to the nature of asymptotic AdS spacetime, it is necessary to consider a regularized WDW patch to avoid the coordinate divergence that arises near the boundary. 

A universal feature of holographic complexity is its linear growth at late times. The rate of this growth (or slope) is shown to be constrained by Lloyd's bound in the AdS-Schwarzchild black hole~\cite{Brown:2015lvg, Engelhardt:2021mju}. Lloyd's bound is proportional to the energy of the boundary states, which corresponds to the mass of the black hole. However, it has been observed that, for certain periods, the growth rate can exceed this bound, violating Lloyd's bound at specific times~\cite{Aguilar-Gutierrez:2023ccv}. 

One of the interesting problems is the exploration of differences between CV and CA proposals, as this may lead to a more precise definition of holographic complexity and its relation to boundary complexities \cite{Iliesiu:2021ari, Yang:2020tna}. For instance, the regularized wormhole length has been proposed as dual to the Krylov state complexity at the boundary~\cite{Rabinovici:2023yex}. To better understand this relationship, it is crucial to test CV and CA proposals in various geometric settings, particularly those with non-trivial boundaries. Notably, CV and CA are found to have different divergent behaviors in the presence of a single boundary~\cite{Chapman:2018bqj, Braccia:2019xxi}. 

This paper aims to investigate the late-time growth rates of CV and CA in AdS$_3$ wedge holography, with two timelike  AdS$_2$ EOW branes that can be either rigid or fluctuating. In the spirit of wedge holography, holographic complexity can be read from both bulk and brane perspectives. To be more general, we will also consider contributions from intrinsic gravity residing on the branes.

This paper is organized as follows: In the next section, we will review the general setup of wedge holography, including the effective action on two EOW branes, the emergence of JT gravity in AdS$_3$, and proposals of CV and CA for wedge holographic complexity. In sections 3 and 4, we will analyze CV and CA proposals for both rigid and fluctuating branes from the perspectives of the bulk and brane. Section 5 will focus on Lloyd's bound and its relation with late-time growth rates of wedge holographic complexities, the general time-dependence of CV and CA, and the effects of fluctuations on these results. Finally, we will summarize our main results and discuss possible directions for future research.

%
\section{Review of previous work}
\subsection{Wedge holography}\label{sec: wedgeholo}
This section reviews the basic setups and concepts of wedge holography and its geometry background.
We will mainly focus on AdS$_3$ wedge holography, where the bulk is asymptotic AdS$_3$ (semi-)classical Einstein gravity with two asymptotic AdS$_2$ EOW branes denoted as $Q_1$ and $Q_2$:
\begin{equation}
    S_{\text{AdS}_3}=-\frac{1}{16\pi G_3} \int d^3x \sqrt{-g_3} \left[ R_{\text{bulk}}+\frac{2}{l^2_3}\right]-\frac{1}{8\pi G_3} \int_{Q_1+Q_2} d^2x \sqrt{-h} (K-T),
\end{equation}
where $R_{\text{bulk}}$ is the AdS$_3$ bulk Ricci scalar, $h$ is the induced metric on the EOW brane, $K$ and $T$ respectively denote the trace of extrinsic curvatures and tensions for branes $Q_1$ and $Q_2$. 
They shall satisfy the Neumann boundary condition consistent with the foliation geometry~\cite{Israel:1966rt, Misner:1973prb, Takayanagi:2011zk, Fujita:2011fp}:
\begin{equation}
    K_{ab}=(K-T)h_{ab}.
\end{equation}

\begin{figure}
\centering
\begin{tikzpicture}
\tzarc[draw=none,fill=green](3.69,0)(130:230:2.61cm);
\tzarc[draw=none,fill=green](0.31,0)(-130:-230:-2.61cm);
\draw[black](2,0) circle (2) ;
\draw[black,dashed](2,0) circle (1.8) ;
\draw[black] (2,2) arc [radius=2.61, start angle=130, end angle=230];
\draw[black] (2,2) arc [radius=-2.61, start angle=-130, end angle=-230];
\draw[-stealth][thick](0.8,0.3) arc [radius=3, start angle=250, end angle=296];
\node[right] at (3,0.3) {$y$};
\node[left] at (1.3,1) {$Q_1$};
\node[right] at (2.7,1) {$Q_2$};
\draw[-stealth][thick](-0.5,0)--(-0.5,1.5);
\node[left] at (-0.5,0.5) {$r$};
\draw[fill=black,black](-0.5,0) circle [radius=0.03];
\draw[-stealth](4.6,0)--(3.8,0);
\node[right] at(4.6,0) {$r=r_{max}$};
\draw[-stealth](2.5,2.5)--(2,1.6);
\node[above] at (2.5,2.5) {AdS$_d$ Wedge};
\draw[fill=blue,blue](2,2) circle [radius=0.05];
\draw[fill=blue,blue](2,-2) circle [radius=0.05];

\draw[-stealth][thick](2.6,-1)--(7,-1);
\draw[black](7.5,-1.5)--(11,-1.5);
\draw[blue,thick](7.5,-1.5)--(7.5,2);
\draw[black](7.5,2)--(11,2);
\draw[blue,thick](11,2)--(11,-1.5);
\draw[black](7.5,2)--(11,-1.5);
\draw[black](7.5,-1.5)--(11,2);
\draw[-stealth](9.25,-2)--(11,-2);
\node[below] at (10,-2) {$r$};
\draw[fill=black,black](9.25,-2) circle [radius=0.03];
\draw[-stealth](11.5,0)--(11.5,2);
\node[right] at (11.5,1) {$t$};
\node[above] at (9.25,2.5) {AdS$_{d-1}$};
\node[below] at (5,-1) {$y$ slice};
\draw[black,dashed] (11,-1.5) arc [radius=5.1, start angle=200, end angle=160];
\draw[black,dashed] (7.5,2) arc [radius=5.1, start angle=20, end angle=-20];
\draw[-stealth](6.4,0)--(7.8,0);

\end{tikzpicture}
\caption{The left figure is a sketch of wedge holography at a constant time slice. Here we show a plot with compactified $y$ coordinate, which is equivalent to the uncompactified case~\cite{Akal:2020wfl}. $Q_1$ and $Q_2$ are two EOW branes with positions at $y_1$ and $y_2$. The blue dots are defects at the asymptotic boundaries. $r_{max}$ is the UV cut-off. The right figure is an AdS$_{d-1}$ slice, where the blue lines are asymptotic boundaries at which the defects are located. } 
\label{fig: wedgeholo}
\end{figure}
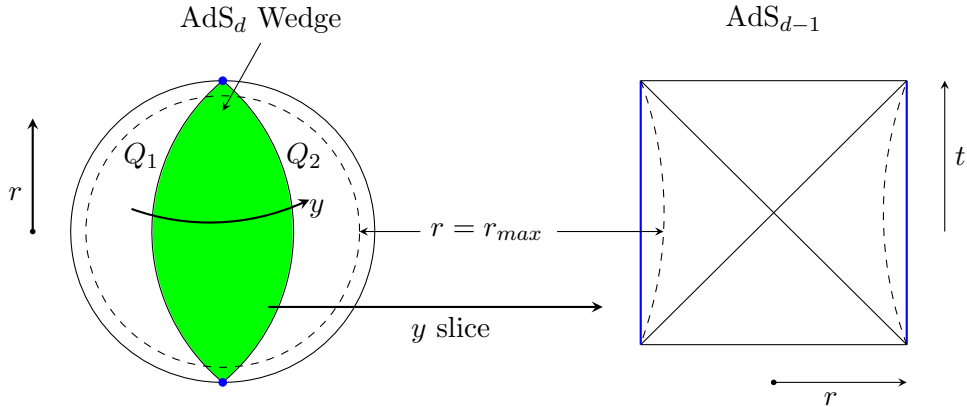

The equation of motion of this action admits AdS$_2$ foliation geometries of AdS$_3$:
\begin{equation}
    ds^2= dy^2+\cosh^2 \left( \frac{y}{l_3} \right) ds^2_{\text{AdS}_2}.
\end{equation}
This geometry well pictures the wedge holography as shown in figure \ref{fig: wedgeholo}, where the above metric corresponds to $d=3$.
We focus particularly on the black string foliation geometry, as it is related to the later discussions on holographic complexity, especially regarding its late-time growth rates. Our discussion will be based on three-dimensional asymptotic AdS spacetime, leading to a geometry background characterized by an AdS$_3$ black string with AdS$_2$ foliations:
\begin{equation}
    ds^2= dy^2+\cosh^2 \left( \frac{y}{l_3} \right)\left( -\frac{r^2-r_h^2}{l_3^2}dt^2+\frac{l_3^2}{r^2-r_h^2}dr^2 \right).
    \label{eq:AdSsclicingBH}
\end{equation}
It is also noteworthy that the AdS$_3$ black string geometry can be obtained through coordinates transformations from AdS$_3$ global patch (see appendix \ref{sec: coordinatesAdS3}).

In this wedge spacetime, it is proposed to be a codimension two holographic system~\cite{Akal:2020wfl, Geng:2020qvw, Bousso:2020kmy, Miao:2020oey}, which can be understood from three equivalent perspectives:

\textbf{Bulk perspective}: Classical Einstein-Hilbert gravity.

\textbf{Brane perspective}: Two codimension-one CFTs live on the branes, minimally coupled with intrinsic Dvali-Gabadadze-Porrati (DGP) gravity on the branes. The two branes are connected via defects located at the asymptotic boundaries. 

\textbf{Boundary perspective}: A codimension-two CFT lives at the shared asymptotic boundary of branes.

Consequently, the complexity of boundary codimension-two CFT has not only just one, but two holographic correspondences. Later, we will primarily focus on CV and CA proposals from bulk and brane perspectives and examine their equivalence in terms of holographic complexity.

\subsection{Tensionless limit and emergent JT gravity}\label{sec: emergentJT}
In \cite{Geng:2022slq, Geng:2022tfc}, it is shown that an emergent JT gravity action arises from wedge holography in the Karch-Randall braneworld with two fluctuating branes in the tensionless limit. Let us start with an AdS$_3$ gravitational action with two boundaries as two Karch-Randall branes. The action takes the form
\begin{eqnarray}
    S_{\text{AdS}_3}&=&-\frac{1}{16\pi G_3} \int dyd^2x \sqrt{-g_3} \left[ R_{\text{bulk}}+\frac{2}{l_3^2}\right]-\frac{1}{8\pi G_3} \int_{Q_1} d^2x \sqrt{-h_1} (K_1-T_1)\nn\\
    &&-\frac{1}{8\pi G_3} \int_{Q_2} d^2x \sqrt{-h_2} (K_2-T_2).
    \label{eq:bulkaction}
\end{eqnarray}

Here, instead of two rigid branes, we consider two fluctuating branes with locations set as $y=y_1+\d \phi_1(r) < 0$ and $y=y_2+\d \phi_2(r) > 0$, where $r$ denotes the radial coordinate living on the brane. 
By solving the Israel junction condition, we can find the branes tensions are
\begin{equation}
    T_1=\tanh \frac{\left\vert y_1 \right\vert}{l_3},\ \  \ T_2=\tanh \frac{y_2}{l_3},
\end{equation}
which are half of the trace of extrinsic curvatures without brane fluctuations. This action admits the foliation geometry as a solution in AdS$_3$ black string as we showed in section \ref{sec: wedgeholo}:
\begin{equation}
ds^2=dy^2+\cosh^2\left(\frac{y}{l_3}\right) g_{2ab}(x) dx^a dx^b,
\label{eq:bulkgeo}
\end{equation}
where $x^{a}=(t,r)$ are the coordinates on the brane, and assume that $\d \phi_1(r), \d \phi_2(r) \ll l_3$, and the induced metric on the brane is $g_{2ab}(y,x) \approx g_{2ab}(x)$ if we only consider lowest graviton mode (low energy limit). We set $g_{\m y}=0$ and $g_{yy}=1$ after fixing the bulk diffeomorphism. We thus have
\begin{equation}
    \sqrt{-g_3}=\cosh \left(\frac{y}{l_3}\right) \sqrt{-g_2}.
\end{equation}
 
Given the bulk geometry (\ref{eq:bulkgeo}), we can write down the bulk Ricci scalar as:
\begin{equation}
R_{\rm{bulk}}=\frac{1}{\cosh^2(y/l_3)}R_{\text{brane}}(x)-\frac{4+2\tanh^2(y/l_3)}{l^2_3}.
\label{eq:bulkR}
\end{equation}
 From now on, we will keep the brane tensions fixed and extrinsic curvatures that depend on the brane's internal fluctuating. If we only keep linear terms of $\d \phi_1(r)$ and $\d \phi_2(r)$, we can then find out the extrinsic curvatures as:
\begin{equation}
K_1=2\tanh\left(\frac{-y_1-\d \phi_1(r)}{l_3}\right), \ \ \ K_2=2\tanh\left(\frac{y_2+\d \phi_2(r)}{l_3}\right).
\label{eq:ecurv}
\end{equation}
We can thus plug (\ref{eq:bulkR}) and (\ref{eq:ecurv}) into the bulk action (\ref{eq:bulkaction}). The effective gravitational integral turns out to be
\begin{eqnarray}
    S_{\text{eff}}=-\frac{1}{16\pi G_3} \int^{y_2+\d\phi_2(r)}_{y_1+\d\phi_1(r)} dy \int dtdr  \sqrt{-g_2}\cosh^2 \left(\frac{y}{l_3}\right)
     \left[ \frac{R_{\text{brane}}(x)}{\cosh^2(y/l_3)}-\frac{2+2\tanh^2(y/l_3)}{l^2_3}\right].\nn\\
\end{eqnarray}

We can then denote $y_2-y_1=\phi_0$ and $\d \phi_2(r)-\d \phi_1(r)=\phi(r)$, where $\phi_0$ is the constant dilaton and $\phi(r)$ is the radion in the context of Randall-Sundrum Braneworld I~\cite{Randall:1999ee} (it is identified as the dynamical dilaton field in two dimensions). We are able to rewrite the effective action as:
\begin{equation}
S_{\rm{eff}}=-\frac{\phi_0}{16\pi G_3} \int d^2 x \sqrt{-g_2} R\lb g_2 \rb -\frac{1}{16\pi G_3} \int d^2x  \sqrt{-g_2}\phi(x)\left( R\lb g_2 \rb+\frac{2}{l_3^2} \right),
\end{equation}
which is the pure JT gravity action, where the first term is the topological term and the second term is the linear dilaton potential term. There will also be dilaton dynamical terms that are irrelevant to our discussion.

Notably, our review of emergent JT gravity above and calculations later depend on the small fluctuation (tensionless limit). It is thus important to discuss the validity of dilaton profiles in our calculations. In the metric we choose (\ref{eq:AdSsclicingBH}), the time-independent dynamical dilaton profile is the solution of the equation of motion, which is written as
\begin{equation}
    \phi(r)=\frac{\phi_h}{r_h}r.
    \label{eq: 2D dilaton}
\end{equation}
At the cut-off boundary $r_{max}$ shown in the figure \ref{fig: wedgeholo}, the dilaton profile behaves as $\phi_b \sim \frac{\phi_h}{r_h} r_{max}$ or $\phi_b \sim \frac{\phi_h}{r_h \e} $ . As approaching the asymptotic boundary $r_{max} \ra \infty$ or $\e \ra 0$, our calculation is valid only when $\frac{\phi_h}{l_3}$ approaching to zero faster than $\e$, namely, it is of order $O(\frac{1}{r^2_{max}})$ or $O(\e^2)$ \cite{Geng:2022slq,Geng:2022tfc}. Practically, in our calculations, we consider $\phi_h \ll \phi_0 \ll l_3$ and $\phi(r_{max}) \ll \phi_0$ which, in the high-dimensional Reissner–Nordström black hole, are understood as the near extremal limit \cite{Maldacena:2016upp}.

\subsection{Brane effective action}\label{sec:Intrinsic brane gravity}
In our setup, we can further consider two EOW branes with intrinsic DGP gravity. Let us only consider the case where the bulk is asymptotically AdS$_3$ and branes are asymptotically AdS$_2$~\cite{Chen:2020uac}. The intrinsic brane action is written as:
\begin{equation}
    S_{\text{brane}}=\frac{1}{16 \pi G_{\text{brane}}} \int d^2 x \sqrt{-g_2} R_{\text{brane}},
\end{equation}
where the metric $g_2$ is the induced metric on the brane and $R_{\text{brane}}$ is the intrinsic curvature scalar only involving induced metric $g_2$. Adding this Einstein-Hilbert term in two dimensions does not change the position of the brane \cite{Chen:2020uac}, namely the Israel junction condition\footnote{This is no longer valid in higher dimensions.}.

The total effective action on the branes should be given as
\begin{equation}
S^Q_{\text{eff}}=S^{Q_1}_{\text{induced}}+S^{Q_2}_{\text{induced}}+S^{Q_1}_{\text{brane}}+S^{Q_2}_{\text{brane}},
\end{equation}
where the induced action is
\begin{equation}
    S^{Q}_{\text{induced}}=S^{Q_1}_{\text{induced}}+S^{Q_2}_{\text{induced}}=S^Q_{\text{gravity}}+S^Q_{\text{CFT}},
\end{equation}
 and $S^{Q}_{\text{CFT}}=-\frac{1}{8\pi G_3} \int_{Q_1} d^2 x \sqrt{-g_2} T^{Q_1}_{\text{brane}}-\frac{1}{8\pi G_3} \int_{Q_2} d^2 x \sqrt{-g_2} T^{Q_2}_{\text{brane}}$, where $T_{\text{brane}}$ is the brane tension that will be specified later, and $S^Q_{\text{gravity}}=-\frac{1}{16\pi G_3} \int^{y_2}_{y_1} dy \int d^2x \sqrt{-g_3} \left[ R_{\text{bulk}}+\frac{2}{l^2_3}\right]-\frac{1}{8\pi G_3} \int_{Q_1+Q_2} d^2x \sqrt{-h} K$ is the action induced on the brane by the bulk gravity.
To find the induced gravitational action~\cite{Chen:2020uac, Geng:2023iqd, Myers:2024zhb} on two branes (we consider branes in a symmetric position, namely, $y_2=-y_1=y_b$), one needs to integrate the bulk action with Gibbons-Hawking-York (GHY) terms over the radial direction (in our case, it is $y$-direction). In the wedge holography setup, two branes play the role of regulator surfaces so that no counter terms are needed. A convenient way of doing so is to consider the Fefferman-Graham expansion of the bulk metric near the brane~\cite{Fefferman:1985, Fefferman:2007rka}. Consequently, the boundary action after integration would only involve terms related to the intrinsic curvature of the boundary metric. The total induced action on two branes can be written as a dilaton $\Phi$ gravity action~\cite{Chen:2020uac, Myers:2024zhb}
\begin{eqnarray}   
    S^Q_{\text{induced}}= \frac{l_3}{16\pi G_3} \int d^2x \sqrt{-g_2} (f(\Phi)+f'(\Phi)(R_{\text{brane}}-\Phi))+\frac{l_3}{8\pi G_3} \int dx \sqrt{-\g} K f'(\Phi),\nn\\
\end{eqnarray}
where
\begin{equation}
    f(\Phi)=\Phi \log \left( \frac{\sqrt{4+2\Phi l^2_3}+2}{\sqrt{4+2\Phi l^2_3}-2}\right)+\frac{1}{T_{\text{brane} } l_3} \left( \Phi+\frac{2}{l^2_{\text{brane}}}\right)+ \sum_{n>1} c_n \left(\Phi+\frac{2}{l^2_{\text{brane}}} \right)^n .
\end{equation}
The on-shell condition from this induced action is easy to find as
\begin{equation}
    \Phi=R_{\text{brane}}=-\frac{2}{l^2_{\text{brane}}}=-\frac{2}{l_3^2}+2T^2_{\text{brane}}.
\end{equation}
After applying the on-shell condition for dilaton $\Phi$, the induced action on the branes can be written as an effective $f(R)$ gravity:
\begin{equation}
    S^Q_{\text{induced}}=\frac{l_3}{16\pi G_3} \int d^2x \sqrt{-g_2}f(R) ,
\end{equation}
where 
\begin{equation}
    f(R)=R\log \left( \frac{\sqrt{4+2R l^2_3}+2}{\sqrt{4+2R l^2_3}-2}\right)+\frac{1}{T_{\text{brane}} l_3} \left( R+\frac{2}{l^2_{\text{brane}}}\right)+ \sum_{n>1} c_n \left(R+\frac{2}{l^2_{\text{brane}}} \right)^n .
\end{equation}
After imposing the on-shell condition for the curvature scalar, the action becomes
\begin{equation}
    S^Q_{\text{induced}}=\frac{l_3}{16\pi G_3} \int_Q \sqrt{-g_2} \log\left( \frac{1+l_3T_{\text{brane}}}{1-l_3T_{\text{brane}}} \right) R_{\text{brane}} d^2x.
\end{equation}
Further adding the brane intrinsic DGP terms, we find that the total action on the branes can be written as an effective Einstein-Hilbert action (we set AdS$_3$ length $l_3=1$ from now on)
\begin{equation}
    S^Q_{\text{eff}}=\frac{1}{16 \pi G_{\text{eff}}}\int_Q \sqrt{-g_2} R_{\text{brane}} d^2x,
    \label{eq:effective action brane}
\end{equation}
where the effective Newton constant on the brane is
\begin{equation}
    \frac{1}{G_{\text{eff}}}=\frac{1}{G_3} \log \left( \frac{1+T_{\text{brane}}}{1-T_{\text{brane}}}\right)+\frac{2}{G_{\text{brane}}},
    \label{eq: effective G on brane}
\end{equation}
and the Ricci scalar on the brane is
\begin{equation}
    R_{\text{brane}}=-2(1-T^2_{\text{brane}}).
    \label{eq: brane ricci}
\end{equation}

\subsection{Wedge holographic complexities}\label{sec:defwedgeHC}
The holographic duality in wedge spacetime is particularly intriguing due to the presence of double duals. This feature offers new perspectives when studying holographic complexity in the context of wedge holography.

In the traditional AdS/(B)CFT framework, holographic complexity is defined as the scalar quantity in the bulk that corresponds to codimension-one state complexity in the boundary. However, in wedge holography, the conformal boundary of bulk space forms a codimension-two defect on the boundary CFT. Therefore, we must consider the codimension-two boundary state complexity, which corresponds not only to bulk scalar quantities but also to scalar quantities on the two branes.

For the CV proposal, the conventional definition is the extremized codimension-one hypersurface anchored to the boundary states:
\begin{equation}
    CV_{B}=\text{max}\left[\frac{V_{\text{bulk}}}{G_N l_N}\right],
\end{equation}
where $V_{\text{bulk}}$ means the volume of the codimension-one hypersurface from the bulk perspective. However, in our setup with an intrinsic DGP term on the brane, inspired by~\cite{Chen:2020uac, Hernandez:2020nem} in which in order to count the holographic entanglement entropy one needs to modify the RT formula to include an additional contribution caused by the DGP term on the brane, the state complexity at the boundary shall receive contributions from two branes in a similar manner. We need to add extremized volume on the branes bounded by the bulk extremal points:
\begin{equation}
CV_{Q}=\text{max}\left[\frac{V_{\text{brane}}}{G_{\text{brane}} l_{\text{brane}}}\right],
\end{equation}
where $V_{\text{brane}}$ is the volume of the codimension-two hypersurface on the brane anchored to the same boundary states.

Since the induced action on two branes can be written effectively as an Einstein-Hilbert action with an effective Newton constant (see section \ref{sec:Intrinsic brane gravity}), the CV proposal can also be written from the brane perspective:
\begin{equation}
\widetilde{CV}_{Q}\footnote{We use symbols with tildes to denote quantities calculated from the brane perspective.}=\text{max}\left[\frac{V_{\text{brane}}}{G_{\text{eff}} l_{\text{brane}}}\right].
\end{equation}
Therefore, the CV proposal with our setup in wedge holography is
\begin{equation}
\boxed{CV_{\text{Wedge}}=CV_{B}+2CV_{Q}=\widetilde{CV}_{Q}.}
\label{eq:WedgeCV}
\end{equation}

For the CA proposal, we need to evaluate four parts: bulk, boundary, joint, and brane contributions from the bulk perspective. The bulk term is given as
\begin{equation}
    A_{\text{bulk}}=\frac{1}{16\pi G_3} \int_{\text{WDW}} d^3x \sqrt{-g} (R-2\L),
\end{equation}
where this term is the bulk on-shell Einstein-Hilbert action evaluated in the WDW patch.
The boundary terms are written as
\begin{equation}
    A_{\text{bdy}}=\frac{\e_K}{8\pi G_3} \int_{\calb_{s}} d^2x \sqrt{|h|} K + \frac{\e_\kappa}{8\pi G_3} \int_{\calb_{n}}d\l d\theta \sqrt{\g} \kappa+\frac{\e_{\Theta}}{8\pi G_3} \int_{\calb_n} d\l dy \sqrt{\g} \Theta \log{( |\Theta|)},
\end{equation}
where the first term is the boundary GHY term for the spacelike surfaces, $h$ is the induced metric on the boundary surface and $K$ is the trace of the extrinsic curvature of the boundary. For the spacelike boundary, $\e_K=+1(-1)$ if the normal vector points into (out of) the WDW patch. The second term is the null boundary term, where $\lambda$ is the affine parameter and $\theta$ is the coordinate that is constant on each generator of the null hypersurface. $\kappa$ is the surface gravity defined to satisfy the geodesic equation: $k^\m \nabla_\m k^\n =\kappa k^\n$, where $k^\m=\frac{dx^\m}{d\lambda}$ is the future-directed null normal vector. And $\e_\kappa=+1(-1)$ according to the WDW patch lying in the future (past) of the null boundary. The third term is the counter term, which is introduced to remove the parametrization dependence of the null generators. $\Theta=\partial_\lambda \log \g$, where $\g$ is the determinant of the metric on the null hypersurface. $\e_{\theta}=+1(-1)$ if the WDW lying in the future (past) of the null hypersurface.

The joint terms that will be considered in this paper are given as
\begin{equation}
    A_{\text{joint}}=\frac{\e_a}{8\pi G_3} \int_{J_n} dx \sqrt{\g} a +\frac{\e_{\eta}}{8\pi G_3} \int_{J_{ts}}dx \sqrt{|\sigma|} \eta,
\end{equation}
where the first term is the null boundary involved joint, $a=\log|k\cdot s|$ for null and timelike joints, $a=\log|k\cdot n|$ for null and spacelike joints, and $a=\log|k\cdot k'/2|$ for null and null joints, where $k$ and $k'$ are null vectors tangent to the null boundary, $s$ is the unit normal vector to the timelike surface, and $n$ is the unit normal vector to the spacelike surface. The sign $\e_a=+1$ if the WDW patch lies at the future (past) of the null boundary and the past (future) of the joint. For all other cases,  $\e_a=-1$. The second term is the joint of timelike and spacelike hypersurfaces, where $\sigma$ is the determinant of the metric living on the joint. $\eta=\log|n+p|\cdot s$, where $s$ is the unit normal vector to the timelike surface, which points outside of the WDW patch; $n$ is the unit normal vector to the spacelike surface, which points to the future direction; $p$ is the unit spatial vector on the spacelike surface and orthogonal to the joint. The sign $\e_\eta=+1$ if $n$ points outside of the WDW patch. Otherwise, $\e_\eta=-1$.

The brane terms are given as
\begin{equation}
    A_{\text{brane}}=\frac{1}{8\pi G_3} \int_{Q\cap \text{WDW}} d^2x \sqrt{-h} (K-T)+\frac{1}{16 \pi G_{\text{brane}}} \int_{Q\cap \text{WDW}} d^2 x \sqrt{-g_2} R_{\text{brane}},
\end{equation}
where we considered brane action from bulk gravity, the CFTs living on the branes, and intrinsic DGP terms.
Therefore, from the bulk perspective, the total on-shell action constrained by the WDW patch in the CA proposal is
\begin{equation}
A_B=A_{\text{bulk}}+A_{\text{bdy}}+ A_{\text{joint}}+A_{\text{brane}},
\end{equation}
where $A_B$ denotes the action from the bulk perspective.

However, due to the nature of wedge holography, the defects at both boundaries are holographically dual to the bulk AdS$_3$ spacetime and two AdS$_2$ branes. Therefore, from the brane perspective, as discussed in section \ref{sec:Intrinsic brane gravity}, the intrinsic action on the brane $Q_1$ and $Q_2$ shall be taken into account as well as the boundaries, joints, and counter terms on the branes:
\begin{equation}
\tilde{A}_{Q}=\tilde{A}_{\text{eff}}+\tilde{A}_{\text{bdy}}+\tilde{A}_{\text{joint}},
\end{equation}
where the first term is the effective action on the two-dimensional brane (\ref{eq:effective action brane}), the other three terms are also on the brane but follow the same rules discussed in the bulk perspective. Here, we denote $\tilde{A}_{Q}$ as the action evaluated in the WDW patch from the brane perspective. 

Therefore, given what we defined above and inspired by \cite{Akal:2020wfl}, the CA proposal in our setup for wedge holography is
\begin{equation}
   \boxed{ A_{\text{Wedge}}=A_{\text{B}}=\tilde{A}_{\text{Q}}.}
    \label{eq:WedgeCA}
\end{equation}

In the next two sections, we will test the late-time growth rates of CV~(\ref{eq:WedgeCV}) and CA~(\ref{eq:WedgeCA}) proposals in AdS$_3$ wedge spacetime with AdS$_2$ branes in the tensionless limit\footnote{Note that when considering higher dimensional geometry and higher curvature theories on branes, the proposals need to be modified, see \cite{Chen:2020uac} and \cite{Hernandez:2020nem}.}. When branes are rigid, we find these proposals are exactly valid. When branes fluctuate, they remain valid in the leading order of fluctuation.

%
\section{Complexity=Volume}
In this section, we test the CV proposal (\ref{eq:WedgeCV}) in AdS$_3$ black string geometry. While similar calculations were done solely from the bulk perspective \cite{Bhattacharya:2023drv}, in this paper, we consider an additional contribution from brane's intrinsic gravity and we will show the matching results from both bulk and brane perspectives for rigid and fluctuating branes within the framework of wedge holography. As reviewed in section \ref{sec: emergentJT}, the results from the CV proposal shall agree with those from JT gravity, as shown in \cite{Bhattacharya:2023drv}. Here, we focus specifically on the late-time growth rate of complexity, leaving the discussion of its saturation value, its relation to Lloyd's bound, and its general time dependence in section \ref{sec: LLoyd and time dependence}.

\subsection{Rigid branes}
\subsubsection{Bulk perspective}

 In AdS$_3$ wedge holography, we consider the bulk black string metric:
\begin{equation}
    ds^2= dy^2+\cosh^2 y\left( -(r^2-r_h^2)d t^2+\frac{1}{r^2-r_h^2}dr^2 \right),
\end{equation}
where the AdS$_3$ length is 1.
Since we consider symmetrically located branes, the induced metrics on two branes share the following metric: 
 \begin{equation}
     ds^2_{Q_1/Q_2}=\cosh^2 y_b\left( -(r^2-r_h^2)d t^2+\frac{1}{r^2-r_h^2}dr^2 \right),
 \end{equation}
 where the AdS$_2$ length on each brane is $\cosh y_b$.

Consider the volume in the bulk enclosed by two rigid branes, which is anchored by boundary states at $t_b$ and $-t_b$:
\begin{eqnarray}
    V_{\text{bulk}}&=&\int_{-y_b}^{y_b } dy \int dr \cosh y \sqrt{-f(r)t '^2+\frac{1}{f(r)}}\nn\\
    &\approx & 2y_b \int dr  \sqrt{-f(r)t '^2+\frac{1}{f(r)}},
\end{eqnarray}
where $f(r)=r^2-r_h^2$, and we take the tensionless limit in the second line. To extremize this volume functional, we defined the Lagrangian as:
\begin{eqnarray}
    L_B=\sqrt{-f(r)t '^2+\frac{1}{f(r)}}.
\end{eqnarray}
Since the Lagrangian does not explicitly depend on $t$, there would be a conserved quantity along the whole hypersurface given as:
\begin{equation}
    P_B=-\frac{\partial L_B}{\partial t'(r)}=-\frac{t' f(r)}{\sqrt{-f(r)t '^2+\frac{1}{f(r)}}}.
\end{equation}
We can solve this to find
\begin{equation}
    t'(r)=\frac{P_B}{f(r) \sqrt{P_B^2 +f(r)}}.
\end{equation}
The boundary time can be expressed as the integral of the above solution from the bulk turning point (extremal point) $r_t$ of the volume functional to the UV cutoff $r_{max}$:
\begin{equation}
    t_b =\int ^{r_{max}}_{r_t} dr \frac{P_B}{f(r) \sqrt{P_B^2 +f(r)}}.
\end{equation}
At this turning point, the time derivative of $r$ should vanish $1/t'(r_t)=0$, which gives us
\begin{equation}
    P_B= \sqrt{r_h^2-r_t^2}.
\end{equation}

For late times $t_b \ra \infty$, the conserved quantity $P_B$ reaches a critical value $P_B^c(\infty)$ which is independent of $r_t$. Therefore, by setting $\frac{\partial P_B}{\partial r_t}=0$, we find the critical value of the conserved quantity:
\begin{equation}
    P_B^c(\infty)=r_h.
\end{equation}
The late-time growth rate of CV from the bulk in the rigid branes case is
\begin{equation}
   \left. \frac{d CV_{B}}{d t_b} \right\vert_{t_b \ra \infty}=\frac{4y_b}{G_3 } P_B^c(\infty)=\frac{2\phi_0 r_h}{G_3 } .
   \label{eq: CVbulk}
\end{equation}
Since in this paper, we consider branes with intrinsic DPG terms, there will be contributions from branes \cite{Hernandez:2020nem}. For each brane, the volume functional in the tensionless limit is given as
\begin{equation}
    V_Q \approx \int dr  \sqrt{-f(r)t '^2+\frac{1}{f(r)}}.
    \label{eq: brane rigid volume}
\end{equation}
Following the analysis in the bulk perspective, it is easy to find the critical value of the conserved quantity at the late time as
\begin{equation}
    P^c_{Q}(\infty)=r_h.
\end{equation}
Therefore, the late-time growth rate of CV from one brane in the rigid branes case is
\begin{equation}
   \left. \frac{d CV_Q}{d t_b} \right\vert_{t_b \ra \infty}=\frac{2}{G_{\text{brane}} } P_{Q}^c(\infty)=\frac{ 2r_h}{G_{\text{brane}} } .
   \label{eq: CVbrane}
\end{equation}
In total, summing over (\ref{eq: CVbulk}) and (\ref{eq: CVbrane}), the late-time growth rate of CV from the bulk perspective is
\begin{equation}
    \left. \frac{d (CV_B+2CV_Q)}{d t_b} \right\vert_{t_b \ra \infty}=\frac{2\phi_0 r_h}{G_3 } + \frac{ 4r_h}{G_{\text{brane}} } .
    \label{eq: CV bulkpers}
\end{equation}

\subsubsection{Brane perspective}
Inspired by \cite{Hernandez:2020nem}, as we discussed in section \ref{sec:defwedgeHC}, considering two AdS$_2$ branes, the CV proposal can be calculated from both bulk and brane perspectives:
\begin{equation}
    CV_{\text{wedge}}=\frac{V_\text{bulk}}{G_{\text{bulk}} l_{\text{bulk}}}+\frac{2V_\text{brane}}{G_{\text{brane}} l_{\text{brane}}}=\frac{V_\text{brane}}{G_{\text{eff}} l_{\text{brane}}},
\end{equation}
the first equal sign is the bulk perspective given by expression (\ref{eq: CV bulkpers}), and the second equal sign is the brane perspective. We can thus verify our result by calculating the CV from the brane perspective. The volume on the brane is still given by the formula (\ref{eq: brane rigid volume}). By computing the effective action on the brane in section \ref{sec:Intrinsic brane gravity}, we know that in the tensionless limit, the effective Newton constant can be written as
\begin{equation}
    \frac{1}{G_{\text{eff}}}=\frac{\phi_0}{G_3}+\frac{2}{G_{\text{brane}}}.
\end{equation}
The brane length $l_{\text{brane}}$ is approximately one in the tensionless limit. It is known from our previous calculations from the bulk perspective and previous work \cite{Stanford:2014jda, Belin:2021bga, Belin:2022xmt, Myers:2024vve} that the late-time growth rate of CV is proportional to the late-time value of the conserved quantity on the codimension-one hypersurface. We, therefore, find the CV from the brane perspective
\begin{equation}
    \left. \frac{d \widetilde{CV}_{Q}}{d t_b} \right\vert_{t_b \ra \infty}=\frac{2}{G_{\text{eff}} } P^c_{Q}(\infty)=\frac{ 2r_h}{G_{\text{eff}} }= \frac{2\phi_0 r_h}{G_3 } + \frac{ 4r_h}{G_{\text{brane}} },
\end{equation}
which exactly matches the result (\ref{eq: CV bulkpers}) from the bulk perspective. This agreement serves as a non-trivial check of the equivalence between bulk and brane perspectives in wedge holography. Although the volumes calculated from each perspective are completely contrasting,  the CV results exhibit a perfect match. However, achieving this agreement is challenging in the fluctuating branes case. Therefore, in the following subsection, we will show that this agreement holds, at least to the leading order in dilatons.

\subsection{Fluctuating branes}\label{sec:CV fluctuaing}
\subsubsection{Bulk perspective}
In the fluctuating branes case, the two branes locations change compared with the rigid branes case. The volume functional $\calv_{\text{bulk}}$\footnote{We use calligraphic letters for quantities in the fluctuating branes case.} is given by
\begin{eqnarray}
    \calv_{\text{bulk}}&=&\int^{y_b+\d\phi_2(r)}_{-y_b+\d\phi_1(r)} dy \int dr \cosh(y) \sqrt{-f(r) t'^2+\frac{1}{f(r)}}\\
    &\approx & \int dr \left(\phi_0+\frac{\phi_h}{r_h}r\right)\sqrt{-f(r) t'^2+\frac{1}{f(r)}},
\end{eqnarray}
where in the second line, we plug in the dilaton solution (\ref{eq: 2D dilaton}).
Similar to our analysis in the rigid branes case, we can define the Lagrangian:
\begin{equation}
    \call_B=\left(\phi_0+\frac{\phi_h}{r_h}r\right)\sqrt{-f(r) t'^2+\frac{1}{f(r)}}.
\end{equation}
We can then find the conserved quantity, which is a constant along the whole hypersurface
\begin{equation}
    \calp_B=-\frac{\partial \call_B}{\partial t'(r)}=-\frac{t' f(r)}{\sqrt{-f(r)t '^2+\frac{1}{f(r)}}}\left(\phi_0+\frac{\phi_h}{r_h}r\right).
\end{equation}
By solving the above equation, we can write the time derivative in terms of the conserved quantity:
\begin{equation}
    t'(r)=-\frac{\calp_B}{f(r) \sqrt{\calp_B^2 +f(r)\left(\phi_0+\frac{\phi_h}{r_h}r\right)^2}}.
    \label{eq: CV fluc tprime}
\end{equation}
The volume functional is extremized at a turning point $r_t$, where the $1/t'(r_t) \ra 0$, which indicates that
\begin{equation}
    \calp_B=\frac{(\phi_0 r_h +r_t \phi_h)}{r_h} \sqrt{r_h^2-r^2_t}.
    \label{eq: CV fluc Pb}
\end{equation}
At the late time, the value of this conserved quantity reaches a critical value $\calp_B^c(\infty)$ which is independent of $r_t$. This condition helps us to find the critical value of the turning point:
\begin{equation}
    \frac{\partial \calp_B}{\partial r_t}=0 \Longrightarrow r^{\text{crit}}_t=\frac{2r_h \phi_h}{\phi_0 +\sqrt{\phi_0^2 +8\phi^2_h}}.
    \label{eq: CV fluc critrt}
\end{equation}
We can then plug this solution into $\calp_B$ and find the critical value
\begin{equation}
    \calp_B^{c}(\infty)=r_h\phi_0+\frac{r_h \phi^2_h}{2\phi_0} +\cdots,
\end{equation}
where `$\cdots$' means higher order terms in $\phi_h$ or $\phi_0$. Therefore, the late-time growth rate of CV from bulk in the fluctuating branes case is
\begin{equation}
   \left. \frac{d \mathcal{CV}_B}{d t_b} \right\vert_{t_b \ra \infty}=\frac{2}{G_3 } \calp^{c}_B(\infty)=\frac{2r_h\phi_0}{ G_3}+\frac{r_h \phi^2_h}{\phi_0 G_3} .
\end{equation}

Next, we evaluate the CV coming from branes due to the intrinsic brane action. We choose two branes located at $y=y_b+\frac{\phi(r)}{2}$ and $y=-y_b-\frac{\phi(r)}{2}$ so that the fluctuations are equally distributed. Considering the linear order in dilatons, the volume functional from one brane is
\begin{eqnarray}
    \calv_{\text{brane}}&=&\int dr \cosh (y_b+\phi(r)/2) \sqrt{-f(r)t '^2+\frac{1}{f(r)}}\\
    &\approx& \int dr \left(1+y_b \frac{\phi_h}{2r_h}r \right)  \sqrt{-f(r)t '^2+\frac{1}{f(r)}},
    \label{eq: fluc brane volfunc}
\end{eqnarray}
where we define the Lagrangian as
\begin{equation}
    \call_Q=\left(1+y_b \frac{\phi_h}{2r_h}r \right)  \sqrt{-f(r)t '^2+\frac{1}{f(r)}}.
\end{equation}
When extremizing the brane volume functional, we want to keep it at the same boundary time as the bulk volume functional. The brane volume is also constrained by the bulk turning point following the proposal in \cite{Hernandez:2020nem}.

By doing so, surprisingly, expanding the volume functional at linear order in dilatons,  we can relate the brane conserved quantity with the bulk one:
\begin{equation}
    \calp_Q=\calp_B \frac{r_h+\phi_0 \phi_h r}{\phi_0 r_h +\phi_h r},
    \label{eq: brane P}
\end{equation}
where $\calp_Q=-\frac{d \call_Q}{d t'(r)}$.
Thus, by using the Leibniz integral rule to the time derivative of (\ref{eq: fluc brane volfunc}), the growth rate of the boundary volume functional can be written as a function of the bulk conserved quantity: 
\begin{equation}
    \frac{d \calv_{\text{brane}}}{d t_b}=\frac{2\call_Q(\calp_B ,r_t)}{t'(r_t)},
\end{equation}
where $\call_Q(\calp_B ,r_t)$ means that the Lagrangian is a function of the turning point $r_t$ and the bulk conserved quantity $\calp_B$ (through equation (\ref{eq: brane P})).
At late times, the growth rate is
\begin{equation}
    \left.\frac{d \calv_{\text{brane}}}{d t_b} \right\vert_{t_b \ra \infty}=\frac{2\call_Q(\calp_B^{c}(\infty) ,r^{\text{crit}}_t)}{t'(r^{\text{crit}}_t)} \approx 2 r_h(1+\phi_h^2).
\end{equation}
At the leading order in dilatons, the late-time growth rate of CV from one brane is
\begin{equation}
    \left.\frac{d \mathcal{CV}_Q}{d t_b} \right\vert_{t_b \ra \infty}=\frac{2 r_h}{G_{\text{brane}}}.
\end{equation}
Therefore, the total CV growth rate at late-time from the bulk perspective is
\begin{equation}
    \left.\frac{d (\mathcal{CV}_{B}+2\mathcal{CV}_Q)}{d t_b} \right\vert_{t_b \ra \infty}=\frac{2r_h\phi_0}{ G_3}+\frac{r_h \phi^2_h}{\phi_0 G_3}+\frac{4 r_h}{G_{\text{brane}}},
    \label{eq: CV fluc}
\end{equation}
where the first two terms agree with the result in \cite{Bhattacharya:2023drv}\footnote{In the boundary theory, the extremal entropy $S_0=\frac{\phi_0}{4G_3}$ corresponds to the number of degrees of freedom $N$ of a certain quantum system. Therefore, this correction term corresponds to the order $\frac{1}{N}$ correction in the large-$N$ limit. We thank Ayan K. Patra for notifying us of this point.}. The second term comes from the fluctuations of branes. By comparing with the result (\ref{eq: CV bulkpers}) from the rigid branes case, we find the brane intrinsic contribution is not changed at the leading order in dilatons. Corrections from fluctuations to the brane intrinsic term turn out to be subleading. This agrees with our findings in CA calculations later.

\subsubsection{Brane perspective}
In this section, we will evaluate the CV from the brane perspective. As pointed out in section \ref{sec:defwedgeHC}, in the spirit of wedge holography, we expect to find the agreement between two perspectives in the fluctuating branes case at least at the leading order in fluctuation. To elaborate on that, we examine the volume from the brane perspective by starting with the CV proposal
\begin{equation}
    \widetilde{\mathcal{CV}}_{Q}=\text{max}\left[\frac{\calv_\text{brane}}{G_{\text{eff}} l_{\text{brane}}} \right],
\end{equation}
where $G_{\text{eff}}$ is the effective Newton constant on the brane with DGP term, which is derived in section \ref{sec:Intrinsic brane gravity}.
Since the effective Newton constant also fluctuates in the fluctuating branes case, we need to extremize the volume including the $log$ term in the effective Newton constant (\ref{eq: effective G on brane}).
Therefore, 
\begin{equation}
    \widetilde{\mathcal{CV}}_{Q}\equiv \text{max}\left[ \calv_{\text{eff}}\right]=\text{max}\left[\frac{1}{G_3}\log\left(\frac{1+T_{\text{brane}}}{1-T_{\text{brane}}}\right) \calv_\text{brane}+\frac{2}{G_{\text{brane}}}\calv_\text{brane}\right].
\end{equation}
At the leading order in dilatons, the effective volume is written as:
\begin{equation}
    \calv_{\text{eff}}\equiv\int dr \call_{\text{eff}}\approx  \int dr \left(\frac{\phi_0+\phi_h}{G_3}+\frac{2+y_b}{G_{\text{brane}}}\phi(r)\right) \sqrt{-f(r)t '^2+\frac{1}{f(r)}} .
\end{equation}
The effective conserved quantity is thus given by:
\begin{equation}
    \calp_{\text{eff}}=-\frac{d \call_{\text{eff}}}{dt'(r)}=\frac{\calp_{B}}{G_3}+\frac{2 \calp_Q}{G_{\text{brane}}},
\end{equation}
where $\calp_{B}$ and $\calp_{Q}$ are given in (\ref{eq: CV fluc Pb}) and (\ref{eq: brane P}).
Given the CV proposal in wedge holography in section \ref{sec:defwedgeHC} and analyses on the turning points, the critical turning point $r^{\text{crit}}_t$ in the brane perspective is the same as in the bulk perspective. We can therefore find 
the overall CV growth rate at late-time from the brane perspective as
\begin{equation}
    \left.\frac{d \widetilde{\mathcal{CV}}_{Q}}{d t_b} \right\vert_{t_b \ra \infty}=2\calp^c_{\text{eff}}(\infty)=2\calp^c_{B}(\infty)+4\calp_Q^c(\infty)=\frac{2r_h\phi_0}{ G_3}+\frac{r_h \phi^2_h}{\phi_0 G_3}+\frac{4 r_h}{G_{\text{brane}}},
\end{equation}
which matches the result from the bulk perspective to the leading order in dilatons.

%
\section{Complexity=Action}
Given the good matches of the CVs from bulk and brane perspectives in the cases of rigid and fluctuating branes, we now evaluate another proposal, ``Complexity=Action''~(CA), in wedge holography for both two cases and two perspectives. One motivation for considering both CA and CV proposals is that they represent fundamentally different notions of holographic complexity, which can lead to distinct complexity results. Interestingly, CA and CV may exhibit similar patterns in lower dimensions; for example, their late-time growth rates are both proportional to the mass of the black hole in JT gravity \cite{Alishahiha:2018swh, Brown:2018bms}. However, their behaviors can differ significantly in higher dimensions~\cite{Chapman:2018bqj, Braccia:2019xxi}. Therefore, we will examine both proposals in AdS$_3$ wedge holography to gain a deeper understanding of CA and CV in more general geometries.  

\subsection{Wheeler-DeWitt Patch}
Based on the definition of CA proposal in section \ref{sec:defwedgeHC}, it is necessary to define the WDW patch in the AdS$_2$ foliation geometry of AdS$_3$ given as:
\begin{equation}
    ds^2=dy^2+\cosh^2y\left[ -f(r)d t^2+\frac{1}{f(r)}dr^2 \right],
\end{equation}
where $y_1<y<y_2$, $f(r)=r^2-r_h^2$, and we set AdS$_3$ length to be 1.

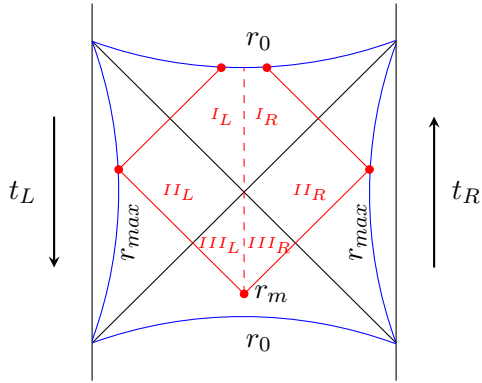
\begin{figure}
\centering
\begin{tikzpicture}
\draw(0,-0.5)--(0,4.5);\draw(4,-0.5)--(4,4.5);
\draw(0,0)--(4,4);
\draw(0,4)--(4,0);

\draw[blue] (0,4) arc [radius=5.8, start angle=20, end angle=-20];
\node[left,rotate=90] at (3.5,2){$r_{max}$};
\node[left,rotate=90] at (0.5,2){$r_{max}$};
\draw[blue] (4,0) arc [radius=5.8, start angle=200, end angle=160];
\draw[blue] (0,0) arc [radius=5.8, start angle=110, end angle=69.7];
\node[left] at (2.5,0){$r_0$};
\draw[blue] (0,4) arc [radius=5.8, start angle=-110, end angle=-69.7];
\node[left] at (2.5,4){$r_0$};
\draw[->][thick](-.5,3)--(-.5,1);
\node[left] at (-.6,2){$t_L$};
\draw[->][thick](4.5,1)--(4.5,3);
\node[right] at (4.6,2){$t_R$};
\draw[fill=red,red](0.35,2.3) circle [radius=0.05];
\draw[fill=red,red](3.65,2.3) circle [radius=0.05];
\draw[red] (0.35,2.3)--(1.7,3.65);
\draw[fill=red,red](1.7,3.65) circle [radius=0.05];
\draw[red] (3.65,2.3)--(2.3,3.65);
\draw[fill=red,red](2.3,3.65) circle [radius=0.05];
\draw[red] (0.35,2.3)--(2,0.65);
\draw[red] (3.65,2.3)--(2,0.65);
\draw[fill=red,red](2,0.65) circle [radius=0.05];
\node[right] at (2,0.65){$r_m$};
\draw[red,dashed] (2,0.65)--(2,3.65);
\node[right,red] at (2,3){$\scriptscriptstyle{I_R}$};
\node[left,red] at (2,3){$\scriptscriptstyle{I_L}$};
\node[right,red] at (2.5,2){$\scriptscriptstyle{II_R}$};
\node[left,red] at (1.5,2){$\scriptscriptstyle{II_L}$};
\node[right,red] at (1.9,1.3){$\scriptscriptstyle{III_R}$};
\node[left,red] at (2.1,1.3){$\scriptscriptstyle{III_L}$};
\end{tikzpicture}
\caption{The region included by red lines is the regularized WDW patch for any constant $y$ direction (on any AdS$_2$ slice). $r_{max}$ is the UV cutoff and $r_0$ is the cutoff behind horizon \cite{Alishahiha:2018swh,Akhavan:2018wla}. $r_m$ is the joint point of two null rays. The boundary of the WDW patch (red solid lines) is approximately null rays in the brane tensionless limit.} 
\label{fig:penroseAdS2}
\end{figure}

The WDW patch is defined as the causal evolution of a Cauchy surface. An equivalent definition is the set of all spacelike geodesics connecting fixed boundary points. It is important to note that the boundaries of the WDW patch consist of null trajectories originating from these boundary states. Consequently, we need to identify null geodesics associated with a specific boundary point, which involves solving for null geodesics in the AdS foliation geometry. To do so, we set $ds^2=0$ for null geodesics, we can find that in any stationary or static spacetime, 
\begin{equation}
    t=g(r,y)+\text{const}\footnote{Similar statements can be found in \cite{Carmi:2016wjl}.},
    \label{eq:ournull}
\end{equation}
where $g(r,y)$ is a function specified by the metric. It is easy to observe that for observers living on any AdS$_2$ slice, the AdS$_3$ null trajectory becomes timelike \cite{Chamblin:1999by}. In the maximally spherical symmetric geometry, this solution simplifies to $t \pm r^*=\text{constant}$, where $r^*$ is the tortoise coordinate. This can provide a constraint to the function $g$ such that
\begin{equation}
    \lim_{y\ra 0} g(r,y) \ra r^*.
\end{equation}
The WDW patch boundaries are light sheets given in (\ref{eq:ournull}), which can be parametrized as:
\begin{equation}
    \calf(t,r,y)=t \pm g(r,y)=\text{constant}.
\end{equation}
In the limit that two branes stay very close to each other (tensionless limit), the AdS$_3$ null rays originating from boundaries or branes can be approximately regarded as AdS$_2$ null rays. Figure \ref{fig:penroseAdS2} shows the regularized WDW patch\footnote{One can also adopt another regularization scheme where the corner of the WDW patch ends on the asymptotic boundaries. However, these two regularization schemes agree once the cutoff is removed $r_{max} \ra \infty$ \cite{Braccia:2019xxi}.} in tensionless limit for each constant $y$ direction.

To further simplify the problem, since we are interested in the late-time growth rate of complexity, it is possible to address it without requiring a specific solution. Notably, we will later show that the exact expression for the null boundaries is not necessary for analyzing the late-time growth rate of the CA conjecture.

\subsection{Rigid branes}
\subsubsection{Bulk perspective}

We begin by evaluating the bulk term in the CA proposal, which involves integrating the on-shell gravitational action in the bulk spacetime. Given that the geometry is symmetric at $r=0$ (the red dashed line in figure \ref{fig:penroseAdS2}), we can separate the spacetime into left and right regions. Each of these regions can further be divided into three areas: I, II, and III, with the boundaries as the future and past horizons. Thus, we can perform the integration in each area and then sum the results. To clarify the coordinate ranges within the WDW patch, we will specify the coordinate limits for each area.
For the right region, 
\begin{eqnarray}
    I_R: &&~  t\in[0, t_R+g(r_{max},y)-g(r,y)],~ r\in[r_h,r_0];\\
    II_R: &&~~ t\in[ t_R-g(r_{max},y)+g(r,y), t_R+g(r_{max},y)-g(r,y)],~ r\in[r_h,r_{max}];\\
    III_R: &&~  t\in[t_R-g(r_{max},y)+g(r,y),0],~ r\in[r_m,r_h].
\end{eqnarray}
For the left region, 
\begin{eqnarray}
    I_L: &&~ t\in[-t_L-g(r_{max},y)+g(r,y),0],~ r\in[r_h,r_0];\\
    II_L: &&~~ t\in[-t_L-g(r_{max},y)+g(r,y),-t_L+g(r_{max},y)-g(r,y)],~ r\in[r_h,r_{max}];\\
    III_L: &&~ t\in[0,-t_L+g(r_{max},y)-g(r,y)],~ r\in[r_m,r_h],
\end{eqnarray}
where the UV cutoff $r_{max}$ and the behind horizon cutoff are related by $r_{max}=\sqrt{\frac{r_h^3}{r_0}}$ \cite{Akhavan:2018wla}, and the joint point $r_m$ is determined by
\begin{equation}
    t_0=2(g(r_{max},y)-g(r_m,y)).
\end{equation}
where $t_0=t_R+t_L$.
We also consider two branes with symmetric locations. Thus, the $y$ coordinate integration region is $ y \in[-y_b,y_b]$. Considering the Ricci scalar in the bulk given in (\ref{eq:bulkR}), we find the action constrained by the WDW patch is
\begin{equation}
    A_{\text{bulk}}=\frac{1}{16\pi G_3} \int^{y_b}_{-y_b}\cosh^2y dy \int_{I+II+III} R_{\text{bulk}} dt dr
\end{equation}
After the integration, since we are interested in the late-time growth rate of complexity, we pick up the time-dependent part: 
\begin{equation}
    A_{\text{bulk}}(t_0)=\frac{t_0}{2\pi G_3} y_b(2r_h-r_0-r_m).
\end{equation}
We observe that the intersection point of two null light-sheets $r_m$ coincides with $r_h$ at late-time. Moreover, we want our final results to be independent of all cutoffs. To do so, we need to take $r_{max} \ra \infty$, which also corresponds to taking $r_0 \ra 0$. We therefore find the finite late-time growth rate of complexity from bulk contribution in the tensionless limit is
\begin{equation}
    \left. \frac{d A_{\text{bulk}}(t_0)}{d t_0}\right\vert_{t_0 \ra \infty}=\frac{r_h\phi_0}{4\pi G_3},
    \label{eq: CArigidbulk-bulk}
\end{equation}
where we identified $2y_b \ra \phi_0$ as in Karch-Randall braneworld.

Next, consider boundary terms in the CA proposal. In total, we have four null boundaries and one spacelike boundary. We can utilize affine parameters to parametrize null boundaries. It is known that the contribution from null boundaries vanishes after affine parametrization. This can be verified by computing the extrinsic curvatures on the null boundaries, which, as expected, turn out to be zero. For counter terms on the null boundaries, as we discussed in section \ref{sec: emergentJT}, we can always fix the diffeomorphism so that $g_{yy}=1$, the counter term $\Theta=k^\m \partial_\m(\log \g)$ vanishes consequently\footnote{One may understand this as a consequence of the broken spherical symmetry. When spherical symmetry is maintained, the induced metric on the null surface corresponds to a $d-2$ sphere in a stationary, spherically symmetric AdS$_d$ spacetime, which leads to non-vanishing counterterms.}.

There is only one spacelike boundary left: $r=r_0$. The induced metric on this  boundary is:
\begin{equation}
    ds^2=dy^2-\cosh^2yf(r_0)d t^2.
\end{equation}
The unit normal vector pointing outward the boundary in the coordinates $(t,r,y)$ is
\begin{equation}
    n_a=\left( 0,-\frac{\cosh y}{\sqrt{f(r_0)}},0 \right).
\end{equation}
Given the ingredients above, one can easily find the trace of extrinsic curvature on this spacelike boundary as
\begin{equation}
    K_s=- \frac{r_0}{\sqrt{f(r_0)}\cosh y}.
\end{equation}
After removing the cutoffs, we can find the late-time growth rate of the action contributed from the boundaries of the WDW patch:
\begin{equation}
   \left. \frac{d A_{\text{bdy}}(t_0)}{d t_0}\right\vert_{t_0 \ra \infty}=0.
    \label{eq: CArigidbulk-bdy}
\end{equation}

Regarding the joint terms, there are three null and null joints, two null and spacelike joints, and two spacelike and timelike joints. It is straightforward to observe that all contributions from joints involving null surfaces are time-independent. Consequently, these joints will not contribute to the late-time growth rate of complexity. We are left with two spacelike and timelike joints. They come from the intersections of spacelike boundary $r=r_0$ and the branes $y=-y_b$ and $y=y_b$. The induced metrics on these joints share the same expression:
\begin{equation}
    ds_{J}^2=-\cosh^2y_b f(r_0) dt^2.
\end{equation}
Following \cite{Lehner:2016vdi}, the general contribution from this joint is written as 
\begin{equation}
    A_{\text{joint}}=\frac{\e_{\eta}}{8\pi G_3} \int dx \sqrt{\left\vert\sigma\right\vert} \eta,
\end{equation}
where $\eta$ is the boost angle between normal vectors defined as
\begin{equation}
    \eta=\log \left\vert(n+p)\cdot s\right\vert,
    \label{eq: boost angle}
\end{equation}
here, $s^\m$ is the unit normal vector to the timelike boundary, $n^\m$ is the unit normal vector to the spacelike boundary, and $p^\m$ is the spatial unit vector orthogonal to the joint in the latter boundary segment. 

In our case, the normal vectors for the timelike boundaries $y=-y_b$ and $y=y_b$ are chosen as
\begin{eqnarray}
     s_{1\m}=(0,0,-1),\ \ \ s_{2\m}=(0,0,1).
\end{eqnarray}

The normal vector for the spacelike boundary $r=r_0$ is
\begin{equation}
    n^{\m}=\left(0,-\frac{\sqrt{f(r_0)}}{\cosh y},0\right).
\end{equation}
Moreover, we find the spatial unit vector orthogonal to the joint and lives in the space-like boundary can be written as
\begin{equation}
    p^\m=(0,0,-1).
\end{equation}
Given all the above, one can calculate the boost angles for two joints by using the formula (\ref{eq: boost angle}). It leads to the vanish of spacelike and timelike joint contributions. However, this happens only in the case of rigid branes. In the section \ref{sec:fluctingbrane} when the branes are fluctuating, there will be non-trivial contributions from joints, which leads to a crucial result on complexity.

The brane contribution to the on-shell action constrained by the WDW patch is generally written as (we include the intrinsic DGP term) 
\begin{equation}
     \frac{1}{8\pi G_3} \int_{Q\cap \text{WDW}} d^2x \sqrt{-h} (K-T)+\frac{1}{16 \pi G_{\text{brane}}} \int_{Q\cap \text{WDW}} d^2 x \sqrt{-h} R_{\text{brane}}.
     \label{eq: braneWDW action}
\end{equation}
Two EOW branes are located at $y=-y_b$ and $y=y_b$ with tensions $T_1=\tanh|-y_b|$ and $T_2=\tanh y_b$. They share the same determinant of metric:
\begin{equation}
    \sqrt{-h_{Q_1/Q_2}}=\cosh^2 y_b.
\end{equation}

Given the induced metric on two branes, one can find the traces of extrinsic curvatures as
\begin{equation}
    K_1=K_2=2\tanh y_b,
\end{equation}
and Ricci scalar on the branes based on equation (\ref{eq: brane ricci}):
\begin{equation}
    R_{\text{brane}}=-2(1-2\tanh^2 y_b).
\end{equation}
One can perform the integral in (\ref{eq: braneWDW action}) and find the late-time growth rate of complexity contributed from branes as:
\begin{eqnarray}
    \left.\frac{d A_{\text{brane}}(t_0)}{d t_0}\right\vert_{t_0 \ra \infty}=-\frac{r_h\phi_0}{8\pi G_3}+\frac{r_h}{4 \pi G_{\text{brane}}}.
    \label{eq: CArigidbulk-brane}
\end{eqnarray}

After summing up (\ref{eq: CArigidbulk-bulk}), (\ref{eq: CArigidbulk-bdy}), and (\ref{eq: CArigidbulk-brane}), we find the total on-shell action growth rate in WDW patch at late-time from the bulk perspective is
\begin{equation}
    \left. \frac{d A_B(t_0)}{d t_0} \right\vert_{t_0 \ra \infty}= \frac{r_h\phi_0}{8\pi G_3}+\frac{r_h}{4\pi G_{\text{brane}}}.
    \label{eq:CAwedgeBulk}
\end{equation}

\subsubsection{Brane perspective}\label{sec:brane perspective}
From the brane perspective, as we discussed in section (\ref{sec:Intrinsic brane gravity}), there is an effective Einstein-Hilbert action on the brane (\ref{eq:effective action brane}) with an effective gravitational constant $G_{\text{eff}}$ given in (\ref{eq: effective G on brane}).
In the rigid branes case, after removing the cutoffs ($r_0 \ra 0, r_{max} \ra \infty$), since brane tensions are constants and AdS$_2$ Einstein gravity is topological, one can compute the on-shell effective action constrained by the WDW patch in AdS$_2$ following \cite{Alishahiha:2018swh}. We find the total time-dependent contribution from two AdS$_2$ branes is:
\begin{equation}
    \tilde{A}_Q(t_0)=\frac{\sqrt{-g_2}r_h t_0}{8\pi G_{\text{eff}}}=\frac{r_h\tau_0}{8\pi}\left(\frac{1}{G_3}\log \left( \frac{1+T_{\text{brane}}}{1-T_{\text{brane}}} \right)+\frac{2}{G_{\text{brane}}} \right).
\end{equation}
In the tensionless limit, we can find the late-time growth rate of on-shell action constrained by the WDW patch from the brane perspective as
    \begin{equation}
    \left.\frac{d \tilde{A}_Q(t_0)}{d t_0} \right\vert_{t_0 \ra \infty}=\frac{r_h\phi_0}{8\pi G_3}+\frac{r_h }{4\pi G_{\text{brane}}}.
    \label{eq:CAwedgeBrane}
\end{equation}

This perfectly matches the result from the bulk perspective as it should be in wedge holography.

\subsection{Fluctuating branes}\label{sec:fluctingbrane}
\subsubsection{Bulk perspective}
For fluctuating branes case, two branes are parametrized as $y=-y_b+\d\phi_1(r)<0$ and $y=y_b+\d\phi_2(r)>0$. 
The WDW patch is still the same as the rigid branes case. The only change is that the integral range in the $y$ direction is changed to be: $-y_b+\d\phi_1(r)<y<y_b+\d\phi_2(r)$. Furthermore, we use the canonical orbifold prescription to project out the fluctuation on one brane~\cite{Geng:2022slq, Geng:2022tfc}\footnote{Orbifolding either $Q_1$ or $Q_2$ is gauge-equivalent. They both lead to pure JT gravity.}. In this paper, we choose brane $Q_1$ to be fixed after orbifolding, which means the range of $y$ becomes: $-y_b<y<y_b+\d\phi(r)$. In the tensionless limit, $\d\phi(r)$ can be expressed as the dilaton solution in two-dimensions:
\begin{equation}
    \d\phi(r)=\phi(r)=\frac{\phi_h}{r_h}r.
\end{equation}

We consider the traces of extrinsic curvatures of two branes to depend on the coordinates living on the branes:
\begin{equation}
    K_1=2\tanh \left\vert-y_b\right\vert, \ \ \ K_2=2\tanh \left(y_b+\phi(r)\right).
\end{equation}
But the tensions of two branes are fixed to be $T_1=\tanh \left\vert-y_b\right\vert$ and $T_2=\tanh y_b$.

The action contribution from bulk is
\begin{equation}
    \cala_{\text{bulk}}=-\frac{1}{16\pi G_3} \int^{y_b+\phi(r)}_{-y_b}\cosh^2y  dy \int_{I+II+III} (4+2\tanh^2 y)dtdr.
\end{equation}
After doing the integration, we extract the time-dependent part:
\begin{eqnarray}
    \cala_{\text{bulk}}(t_0)&=&  -\frac{t_0}{16\pi G_3} \left[ \frac{\phi_h}{2r_h}(r^2_m+r^2_0-2r^2_h) +(2y_b + 6\cosh y_b \sinh y_b)(r_m+r_0-2r_h)\right.\nn\\
 &&\left.+ \frac{3\phi_h}{2r_h} (r^2_m+r^2_0-2r^2_h) (\cosh^2 y_b +\sinh^2 y_b)\right].
\end{eqnarray}
By taking the tensionless limit, the late-time growth rate from bulk contribution is:
\begin{equation}
    \left.\frac{d \cala_{\text{bulk}}(t_0)}{d t_0} \right\vert_{t_0 \ra \infty}= \frac{\phi_0 r_h}{4\pi G_3}+\frac{\phi_h r_h}{8\pi G_3}+\frac{3\phi_h\phi^2_0 r_h}{128\pi G_3}.
    \label{eq: fluc bulkpers bulk}
\end{equation} 

Similar to the rigid branes case, all null-involved boundaries do not contribute to the late-time growth rate of complexity. 
For the spacelike boundary $r=r_0$, the trace of extrinsic curvature is 
\begin{equation}
    K_s=- \frac{r_0}{\cosh y \sqrt{f(r_0)}}.
\end{equation}

The boundary contribution is
\begin{eqnarray}
    \cala_{\text{bdy}}=-\frac{1}{8\pi G_3} \int dt dy \sqrt{|h|} K_s.
\end{eqnarray}
We can find the time-dependent part as
\begin{eqnarray}
     \cala_{\text{bdy}}(t_0)=\frac{r_0 t_0}{8\pi G_3 } \left( \phi_0 +\frac{\phi_h}{r_h}r_0  \right).
\end{eqnarray}
After removing the cutoffs, the late-time growth rate is then
\begin{equation}
   \left. \frac{d  \cala_{\text{bdy}}(t_0)}{d t_0} \right\vert_{t_0 \ra \infty}=0.
   \label{eq: fluc bulkpers bdy}
\end{equation}

For joints, similar to the rigid branes case, those involving the null hypersurfaces do not contribute to the late-time behavior of complexity. However, unlike the rigid branes case, the spacelike and timelike joints will have non-trivial contributions.

The unit normal vector for spacelike boundary $r=r_0$ is
\begin{equation}
    n^{\m}=\left(0,-\frac{\sqrt{f(r_0)}}{\cosh y},0\right).
\end{equation}

The normal vector for the timelike boundary $y_1=-y_b$ is
\begin{eqnarray}
     s_{1\m}=(0,0,-1).
\end{eqnarray}

For timelike boundary $y_2=y_b+\phi(r)$, the unit normal vector is given by:
\begin{equation}
    s_{2\m}
    = \frac{1}{\sqrt{1+\frac{\phi^2_h f(r)}{r^2_h \cosh^2 y_2}}} \left( 0, \frac{\phi_h}{r_h},1 \right).
\end{equation}

The spatial unit vector orthogonal to the joint and living in the spacelike boundary is
\begin{equation}
    p^\m=(0,0,-1).
\end{equation}

Therefore, following the action manual in \cite{Lehner:2016vdi}, the joint contribution is
\begin{equation}
    \cala_{\text{joint}}= \frac{1}{8\pi G_3} \int_{J_{ts}} dt \cosh(y_b+\phi(r_0)) \sqrt{f(r_0)} \log \left( \frac{\frac{\sqrt{f(r_0)}\phi'(r_0)}{\cosh(y_b+\phi(r_0))}+1 }{\sqrt{1+\frac{ \phi'^2(r_0)f(r_0)}{\cosh^2(y_b+\phi(r_0))}}}\right),
\end{equation}
where $\e_{\eta}=+1$ since $n^\m$ is pointing out of the WDW patch.
The late-time growth rate after removing the cutoffs is
\begin{equation}
    \left.\frac{d  \cala_{\text{joint}}(t_0)}{d t_0} \right\vert_{t_0 \ra \infty}
    \approx  \frac{r_h }{16\pi G_3} \log\left(\frac{1+\phi_h}{1-\phi_h} \right).
    \label{eq: fluc bulkpers joint}
\end{equation}

Next, we consider contributions from two branes. As we discussed earlier in this section, after orbifolding, we keep brane $Q_1$ fixed and brane $Q_2$ fluctuating.
For brane $Q_1$, the analysis is the same as that of the rigid branes case. We have:
\begin{eqnarray}
    \cala_{Q_1}
    &=&\frac{1}{8\pi G_3} \int_{I+II+III} dt dr \sinh y_b \cosh y_b.
\end{eqnarray}
The time-dependent part in the tensionless limit is:
\begin{equation}
    \cala_{Q_1}(t_0)=\frac{t_0}{8\pi G_3} y_b(r_m+r_0-2r_h).
\end{equation}
For brane $Q_2$, since the trace of the extrinsic curvature and brane tension are
\begin{equation}
K_2=2\tanh (y_2+\phi(r)),\ \ \ T_2=\tanh y_2.
\end{equation}
Then, the action from brane $Q_2$ up to linear order in dilatons is
\begin{eqnarray}
    \cala_{Q_2}
    &=&\frac{1}{8\pi G_3} \int_{I+II+III} dt dr (\cosh^2 (y_2+\phi(r)))(2\tanh (y_2+\phi(r))-\tanh y_2)
\end{eqnarray}
The time-dependent part in the tensionless limit is
\begin{equation}
    \cala_{Q_2}(t_0)
 = \frac{t_0}{8\pi G_3} (\cosh y_b \sinh y_b)(r_m+r_0-2r_h)+ \frac{t_0}{8\pi G_3} \frac{\phi_h}{r_h}(r^2_m+r^2_0-2r_h^2).
\end{equation}
Summing over contributions from two branes, we find the late-time growth rate is
\begin{equation}
    \left.\frac{d  \cala_Q(t_0)}{d t_0} \right\vert_{t_0 \ra \infty}=-\frac{\phi_0 r_h}{8\pi G_3}-\frac{\phi_h r_h}{8\pi G_3}.
    \label{eq: fluc bulkpers brane}
\end{equation}

At last, we consider contributions from brane intrinsic DGP terms:
\begin{eqnarray}
   \cala^{\text{Intrinsic}}_Q(t_0) &=&-\frac{2}{16 \pi G_{\text{brane}}} \int_{Q_1\cap \text{WDW}} dt dr \cosh^2 y_b (1-\tanh^2 y_b)\nn\\
    &&- \frac{2}{16 \pi G_{\text{brane}}} \int_{Q_2\cap \text{WDW}} dt dr \cosh^2 (y_b +\phi(r))(1-\tanh^2(y_b+\phi(r)))\nn\\
    &=&-\frac{t_0}{4 \pi G_{\text{brane}}} (r_m+r_0-2r_h).
\end{eqnarray}
The late-time growth rate after removing the cutoffs is
\begin{equation}
    \left.\frac{d  \cala^{\text{Intrinsic}}_Q(t_0)}{d t_0} \right\vert_{t_0 \ra \infty}=\frac{r_h}{4 \pi G_{\text{brane}}}.
    \label{eq: fluc bulkpers intrinsic}
\end{equation}

In total, summing over (\ref{eq: fluc bulkpers bulk}), (\ref{eq: fluc bulkpers bdy}), (\ref{eq: fluc bulkpers joint}), (\ref{eq: fluc bulkpers brane}) and (\ref{eq: fluc bulkpers intrinsic}), the late-time growth rate of the on-shell action constrained by the WDW patch in the fluctuating branes case is
\begin{equation}
      \left.\frac{d\cala_B}{dt_0}\right\vert_{t_0 \ra \infty}
      \approx \frac{\phi_0 r_h}{8\pi G_3}+\frac{r_h\phi_h}{8\pi G_3} +\frac{r_h}{4 \pi G_{\text{brane}}},
     \label{eq:CAfluctuatingBulk}
\end{equation}
where for the joint term, we can restore the AdS$_3$ length through dimensional analysis: $\log\left(\frac{1+\phi_h/l_3}{1-\phi_h/l_3} \right)$, assuming the dilaton $\phi_0,\phi_h \ll l_3$ (as we discussed in section \ref{sec: emergentJT}), we can get the result by Taylor expanding the logarithm term and ignoring the higher-order corrections.

In the result above, the first term in the last line gives exactly the holographic complexity in JT gravity as expected. Known the Hawking temperature $T=\frac{|f'(r_h)|}{4\pi}=\frac{r_h}{2\pi}$ and Bekenstein-Hawking entropy $S_0=\frac{\phi_0}{4G_3}$, the first term is proportional to $S_0 T$ which agrees with the known result of holographic complexity in JT gravity~\cite{Brown:2018bms, Alishahiha:2018swh}. Beyond that, compared with the rigid branes case, we find an extra term in the fluctuating branes case, which is a constant term (independent of $\phi_0$).

\subsubsection{Brane perspective}

In this section, we examine the CA proposal from the brane perspective while keeping one brane fixed and one brane fluctuating, which is intended to verify our CA result in fluctuating branes case from the bulk perspective. Similar to our analysis in section \ref{sec:brane perspective}, we set the positions of branes symmetrically but only make the brane $Q_2$ fluctuating: brane $Q_1$ at $y=-y_b$ and brane $Q_2$ at $y=y_b+\phi(r)$. 

We expect the brane effective Newton constant to be dynamical in $r$-direction, which comes from a small fluctuation of the brane. However, this $r$-dependent correction turns out to be subleading. Thus, in the leading order, we keep both effective Newton constants on two branes independent of coordinates and they are \cite{Geng:2022slq,Geng:2022tfc}:
\begin{equation}
    \frac{1}{G^{Q_1}_{\text{eff}}}=\frac{y_b}{G_3}+\frac{1}{G_{\text{brane}}}, \ \ \ \frac{1}{G^{Q_2}_{\text{eff}}}=\frac{y_b+\phi_h}{G_3}+\frac{1}{G_{\text{brane}}},
\end{equation}

We can now evaluate the bulk contributions on brane $Q_1$ and $Q_2$ as 
\begin{equation}
    \tilde{\cala}^{\text{bulk}}_{Q_1}(t_0)=\tilde{\cala}^{\text{bulk}}_{Q_2}(t_0) =\frac{r_h t_0 }{8 \pi G^{Q_2}_{\text{eff}}}.
\end{equation}
Therefore, the total bulk contribution to the late-time growth rate of complexity from two branes in the tensionless limit after removing the cutoffs is:
\begin{equation}
    \left.\frac{d \tilde{\cala}^{\text{bulk}}_{Q}(t_0)}{d t_0}\right\vert_{t_0 \ra \infty}=\frac{r_h}{8\pi }\left(\frac{1}{G^{Q_1}_{\text{eff}}} +\frac{1}{G^{Q_2}_{\text{eff}}}\right)=\frac{\phi_0 r_h}{8\pi G_3}+\frac{\phi_h r_h}{8\pi G_3}+\frac{r_h}{4\pi G_{\text{brane}}}.
\end{equation}

Next, we move on to consider boundary terms. Similarly, the only term that has a non-trivial contribution to the late-time growth rate of complexity is the spacelike boundary $r=r_0$.
The induced metric on this boundary is
\begin{equation}
    ds^2_{r=r_0}=-\cosh^2 (y_b+\phi(r_0))f(r_0)dt^2.
\end{equation}
The unit normal vector of this spacelike boundary that is pointing inside the WDW patch is
\begin{equation}
    n_a=\frac{\cosh (y_b+\phi(r_0))}{\sqrt{f(r_0)}} (0,-1).
\end{equation}

The trace of extrinsic curvature can be derived as 
\begin{eqnarray}
    K_s=-\frac{f'(r_0)}{\sqrt{f(r_0)}} -2\sqrt{f(r_0)}\phi'(r_0)\tanh (y_b+\phi(r_0)).
\end{eqnarray}
The time-dependent part of boundary contribution is
\begin{equation}
   \tilde{\cala}_{\text{bdy}}(t_0)
   =\frac{t_0}{8\pi G_{\text{eff}}} \left( \cosh (y_b+\phi(r_0))f'(r_0)+2 f(r_0)\phi'(r_0)\sinh (y_b+\phi(r_0)) \right)
\end{equation}
After removing the cutoffs and taking the tensionless limit, the boundary contribution is
\begin{equation}
    \tilde{\cala}_{\text{bdy}}(t_0)=-\frac{r_h\phi_0\phi_h t_0}{8\pi G_{\text{eff}}}
\end{equation}

The boundary contribution from brane $Q_1$ vanishes after removing the cutoffs. We find that all the surviving boundary contributions are in higher order in $\phi_0$ and $\phi_h$.

In summary, at leading order in dilaton, the late-time growth rate from the brane perspective is
    \begin{equation}
    \left.\frac{d \tilde{\cala}_Q}{dt_0} \right\vert_{t_0 \ra \infty}=\frac{\phi_0 r_h}{8\pi G_3}+\frac{\phi_h r_h}{8\pi G_3}+\frac{r_h}{4\pi G_{\text{brane}}},
    \label{eq:CAfluctuatingBrane}
\end{equation}
which matches the result from the bulk perspective. The term proportional to $\phi_0$ arises from the case without fluctuation, while the term proportional to $\phi_h r_h$ represents the first-order correction resulting from brane fluctuations. According to the definition of constant dilaton $\phi_0=y_2-y_1$ in the Karch-Randall braneworld, this correction term is independent of the positions of EOW branes, to which we refer as a constant term. From the bulk perspective, we can trace this constant term back to the joint term of the two branes. Specifically, this non-trivial constant term shows up due to edge modes in the corners of the wedge spacetime, similar to the edge modes effects discussed in \cite{Takayanagi:2019tvn, Akal:2020wfl}, which contribute to the late-time growth rate of complexity and vanishes once closing the wedge (there still would be non-zero action or complexity left). Interestingly, this term shows up in the bulk term from the brane perspective, rather than the boundary term which is of higher order in dilatons. Correspondingly, in CV results, we also see such an additional term proportional to $\frac{\phi_h^2}{\phi_0}$, which shares the same asymptotic behavior as the CA result when closing the wedge.

Moreover, by comparing the late-time growth rates of complexity from CA (\ref{eq:CAfluctuatingBrane}) and CV (\ref{eq: CV fluc}) in the fluctuating branes case, we observe different effects of fluctuations to two proposals. Based on our setup, the CV proposal is more sensitive to brane fluctuations compared with the CA proposal, resulting in distinct saturation behaviors as fluctuations increase (discussed further in the next section).  This sensitivity to geometric details may reveal a unique feature of the CV proposal, providing a non-trivial distinctive insight into the dual of holographic complexity.

%
\section{Lloyd's bound and time dependence of holographic complexity}\label{sec: LLoyd and time dependence}
In this section, we discuss Lloyd's bound on the rate of growth for holographic complexity in AdS$_3$ wedge spacetime. Inspired by the Heisenberg uncertainty principle relating time and energy: $\D E \D t \geq \hbar$, which can be interpreted to mean that the information of a quantum state, represented by $\D E$, takes at least $\D t$ time to evolve to an orthogonal state, Lloyd's bound was originally proposed as a limitation on the speed of information spreading among qubits in a quantum physical system \cite{lloyd2000ultimate}:
\begin{equation}
    \frac{1}{\D t} \leq \frac{2E}{\pi}.
\end{equation}
Later, it was conjectured a similar bound for holographic complexities~\cite{Brown:2015lvg}. The energy of a quantum state is holographically dual to the mass $M$ of the black hole. This bound indicates that the holographic complexity for black holes with spherical symmetry grows linearly at late times, with the growth rate bounded by:
\begin{equation}
    \frac{d CA}{dt} \lesssim \frac{2M}{\pi},
\end{equation}
where we use $\lesssim$ since we omit a constant factor that depends on a specific system.
This bound can be tightened if additional conserved charges, such as the electric charge $Q$ or angular momentum $J$, are present. Lloyd's bound was proven holographically through the positive complexity theorem, assuming a weak curvature condition in the spacetime with spherical symmetry~\cite{Engelhardt:2021mju}. 

In JT gravity, the large black hole has the following relation with the temperature and entropy \cite{Brown:2015lvg}:
\begin{equation}
    TS=2M.
\end{equation}
Therefore, in our setup, the black string mass in the bulk and the black hole mass on the brane are given by 
\begin{equation}
    M_{BS} = \frac{\phi_0 r_h}{16\pi G_3},\ \ \ M_{BH}= \frac{r_h}{16\pi G_{\text{brane}}}.
\end{equation}
For the CA proposal, in the case when two branes are rigid, the growth rate of CA (\ref{eq:CAwedgeBulk}) at
late times saturates the bound:
\begin{equation}
    \left.\frac{d CA}{d t_0} \right\vert_{t_0 \ra \infty}=\frac{2M_{\text{BS}}}{\pi}+\frac{4 M_{\text{BH}} }{\pi}.
\end{equation}
When the two branes fluctuate, at leading order in dilatons, the mass of the black string is changed to be
\begin{equation}
\calm_{\text{BS}}\approx \frac{(\phi_0+\phi_h)r_h}{16\pi G_3}.
\end{equation}
The black hole mass on AdS$_2$ branes remains the same in the low energy limit as discussed in section \ref{sec: emergentJT}.
From equation (\ref{eq:CAfluctuatingBulk}), we can see that the late-time growth rate from the CA proposal still saturates the Lloyd's bound:
\begin{equation}
    \left.\frac{d \mathcal{CA}}{d t_0} \right\vert_{t_0 \ra \infty}=\frac{2\calm_{\text{BS}}}{\pi}+\frac{4 M_{\text{BH}} }{\pi}.
\end{equation}

As for the CV proposal, the late-time growth rate in the rigid branes case (\ref{eq: CV bulkpers}) saturates at the value:
\begin{equation}
    \left. \frac{d CV}{d t_b} \right\vert_{t_b \ra \infty}=32\pi M_{\text{BS}} + 64\pi M_{\text{BH}}.
\end{equation}
In the fluctuating branes case, the CV proposal result (\ref{eq: CV fluc}) is always below the bound:
\begin{equation}
    \left. \frac{d \mathcal{CV}}{d t_b} \right\vert_{t_b \ra \infty}<32\pi \calm_{\text{BS}} +  64\pi M_{\text{BH}},
\end{equation}
where the term $\frac{\phi^2_h}{\phi_0}$ coming from fluctuations in CV proposal result is always smaller than $\phi_h$ as discussed in section \ref{sec: emergentJT}. 

\begin{figure}[h!]
    \centering
    \subfloat[]{
    \centering
         \includegraphics[width=0.48\textwidth]{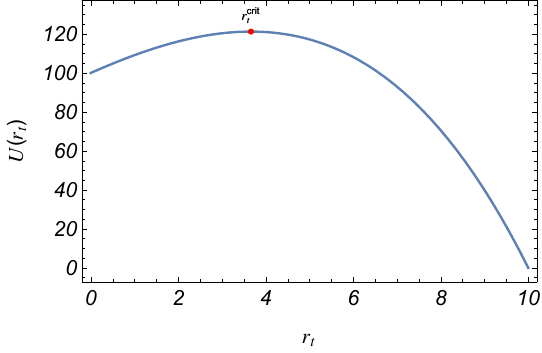}}
    \hfill
    \subfloat[]{
    \centering
         \includegraphics[width=0.48\textwidth]{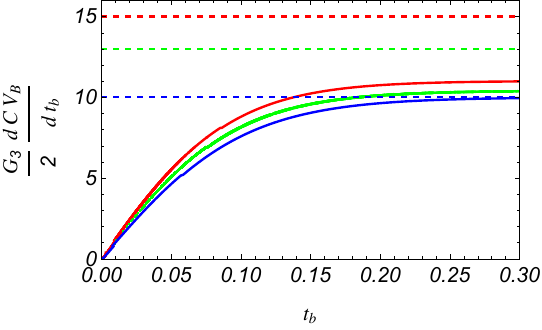}}
    \caption{The left figure is a numerical plot of the effective potential $U(r_t)$ (\ref{eq: eff potential}) with parameters chosen as $\phi_0=1$, $\phi_h=0.5$, $r_h=10$. The figure on the right is a numerical plot of time-dependence of the growth rate of the bulk contributed $\mathcal{CV}_B$ with parameters chosen as $\phi_0=1$ and $r_h=10$. The solid and dashed curves are the growth rate and Lloyd's bound with $\phi_h=0$ (blue), $\phi_h=0.3$ (green), $\phi_h=0.5$ (red).}
    \label{fig:potential and time dependence}
\end{figure}

Furthermore, we can consider the behaviors of CV and CA approaching the saturation value. For the CA proposal, it is straightforward to observe from (\ref{eq:CAwedgeBulk}) that due to the factor $2r_h-r_m$, where $r_m$ approaching $r_h$ from zero,  the growth rate of CA approaches the bound from above, which is valid for both rigid and fluctuating branes cases. As for the CV proposal, in section \ref{sec:CV fluctuaing}, we are able to pick up the saturation value without examining the complexity and its growth rate in detail. However, to find the behavior of complexity growth rate before late times, it is necessary to consider the conserved quantity $\calp_B$ before the critical minimal radius (\ref{eq: CV fluc critrt}). Given equations (\ref{eq: CV fluc tprime}) and (\ref{eq: CV fluc Pb}), the extremality problem can be recast by an effective potential $U(r)$ \cite{Chapman:2018lsv, Carmi:2017jqz, Belin:2021bga, Myers:2024vve}, which is defined through:
\begin{equation}
    \dot{r}^2+U(r)=\calp_B^2.
\end{equation}
This describes a classical particle moving in the potential $U(r)$. The turning point $r_t$ of this classical particle corresponds to the extremal radius point of the volume functional, which satisfies
\begin{equation}
    \dot{r}\vert_{r=r_t}=0.
\end{equation}
The effective potential at the turning point is given as
\begin{equation}
    U(r_t)=\calp_B^2(r_t)=\frac{(\phi_0 r_h+\phi_h r_t)^2}{r^2_h}f(r_t).
    \label{eq: eff potential}
\end{equation}
At late times, this extremal point reaches the critical value (\ref{eq: CV fluc critrt}), which is inside the horizon as shown in the left plot of figure \ref{fig:potential and time dependence}.
To study the behavior of the growth rate approaching the saturation value, we can do the expansion up to the second order of the growth rate around the critical value of radius $r^{\text{crit}}_t$ \cite{Carmi:2017jqz}:
\begin{equation}
    \frac{d \mathcal{CV_{B}}}{dt_b}=\frac{2}{G_3} \left( \sqrt{U(r^{\text{crit}}_t)} +\frac{U''(r^{\text{crit}}_t)(r_t-r^{\text{crit}}_t)^2}{4 \sqrt{U(r^{\text{crit}}_t)}}\right),
    \label{eq: CV expansion}
\end{equation}
where at the critical point the first-order derivative vanishes. The second term $U''(r^{\text{crit}}_t)$ in (\ref{eq: CV expansion}) is always negative since this critical point is a maximum of $U(r_t)$. Thus, we conclude that the growth rate from the CV proposal saturates from below in both cases. This conclusion is further supported by the general time dependence of the CV growth rate in the right plot in figure \ref{fig:potential and time dependence}. As observed, when two branes are rigid, the growth rate saturates the Lloyd's bound from below. However, as fluctuations increase, Lloyd's bound becomes gradually larger than the saturation value.

%
\section{Summary and Discussions}

In this paper, we have explored CV and CA proposals within the framework of wedge holography from both bulk and brane perspectives. From the brane perspective, it is known that the on-shell action on the branes can be effectively expressed as an Einstein-Hilbert action with an effective Newton constant~\cite{Chen:2020uac, Hernandez:2020nem, Myers:2024zhb}. We focus on the limit where two branes are situated close to one another, specifically the tensionless limit. In this limit, it is known that the AdS$_3$ wedge Einstein gravity reduces to JT gravity~\cite{Geng:2022tfc, Geng:2022slq}.  Furthermore, We consider two EOW branes with intrinsic DGP gravity. In this geometric and theoretical setup, we proceed to study holographic complexities for both rigid and fluctuating branes cases and evaluate CV and CA from both bulk and brane perspectives in the spirit of wedge holography.

In the CV proposal, while keeping branes rigid, we find the late-time growth rate of complexity precisely reproduces the result from JT gravity, augmented by the brane intrinsic contributions. We obtain consistent results from both bulk and brane perspectives. However, when the branes fluctuate, the late-time growth rate includes an additional correction term that behaves like $\frac{\phi^2_h}{\phi_0}$, which matches with previous work \cite{Bhattacharya:2023drv}. Results from both perspectives agree on the leading order in dilaton fields.

In the CA proposal, we considered cutoffs at both the UV and behind the horizon to regularize the on-shell action~\cite{Akhavan:2018wla}. The CA late-time growth rate remains finite after removing these cutoffs. In the case of rigid branes, we find an exact match between our CA result and that from JT gravity, along with the brane intrinsic contributions. However, in the fluctuating branes case, we observe an extra constant correction that is independent of $\phi_0$. This correction comes from the joint contribution from the bulk perspective and the bulk contribution from the brane perspective. This non-trivial constant term is believed to be related to the edge modes of gravity since it is contributed mainly by the joint (corner) term. The correction term arising from fluctuations further indicates that the CV proposal is more sensitive to geometric details than the CA proposal, which could be a distinctive feature of holographic complexity and may serve as a potential trace for identifying their duals in the boundary.

Beyond our calculations on the late-time growth rates, we further analyzed these results in relation to Lloyd's bound at late times and their general time dependence. In the CA proposal, growth rates reach Lloyd's bound for both rigid and fluctuating branes cases, with the growth rates approaching the saturation value from above. For the CV proposal, growth rates saturate the bound in the rigid branes case. However, as the brane fluctuations increase, a gap between Lloyd's bound and the growth rates widens, with the bound increasingly exceeding the growth rates. Furthermore, in both cases, the growth rates from the CV proposal consistently approach the saturation value from below.

Lastly, we propose several interesting directions for future research. One natural generalization is to explore holographic complexity for higher-curvature gravity~\cite{Jiang:2023jti, Wang:2023noo, Mandal:2022ztj, Ghodsi:2020qqb, Alishahiha:2017hwg, Ding:2018ibq, Jiang:2018sqj, Cano:2018aqi, An:2018dbz} in the context of (higher-dimensional)wedge holography. In this paper, we focused on asymptotically AdS braneworlds; however, holographic complexity has been studied in dS braneworlds~\cite{Aguilar-Gutierrez:2023tic}. It is known that by tuning the brane tension, one can achieve asymptotically AdS, dS, or flat braneworlds. An intriguing direction would be to investigate the transition of holographic complexity across these different braneworlds by setting the tension at a critical value~\cite{Myers:2024zhb}. Additionally, examining the boundary correspondence of wedge holographic complexity presents another promising avenue for exploration. It is especially interesting to consider the SYK model, in which we may find the fingerprint of the extra constant contribution of complexity in the fluctuating branes case.

%

\acknowledgments
We thank Byoungjoon Ahn, Sang-Eon Bak, Hugo Camargo, Viktor Jahnke, Hyun-Sik Jeong, Kuntal Pal, and Bartosz Pyszkowski for useful discussions. We especially thank Shan-Ming Ruan for inspiring discussions and patient explanations. We are also grateful to Arpan Bhattacharyya, Hao Geng, and Ayan Patra for the correspondence; Byoungjoon Ahn, Hugo Camargo, Viktor Jahnke, Kuntal Pal, and Shan-Ming Ruan for comments on the manuscript. 
 This work was supported by the Basic Science Research Program through the National Research
Foundation of Korea (NRF) funded by the Ministry of Science, ICT \& Future Planning (NRF2021R1A2C1006791) and
the Al-based GIST Research Scientist Project grant funded by
the GIST in 2024. KYK was also supported by the Creation
of the Quantum Information Science R\&D Ecosystem (Grant
No.2022M3H3A106307411) through the National Research
Foundation of Korea (NRF) funded by the Korean government
(Ministry of Science and ICT).
We thank the organizers of the School of Frontiers in Hydrodynamics, Effective Field Theory and Holographic Duality at Jilin University where part of this work was done.
This research was partially supported by the Asia Pacific Center for Theoretical Physics (APCTP) for participating in the APCTP Focus Program, `Entanglement, Large N, and Black Hole 2024'. Yichao Fu is the first author of this paper.
%
\appendix

%
\section{$\text{AdS}_3$ black string geometry}\label{sec: coordinatesAdS3}
In this section, we show how to get the foliation geometry (\ref{eq:AdSsclicingBH}) in the main text by coordinate transformations. To do so, let us start with the \Poincare patch in AdS$_3$ space:
\begin{equation}
	ds^2= \frac{l^2_3}{z^2} \left(  -dt^2+dz^2+dw^2\right),
\end{equation}
where $z$ is the radial coordinate and the asymptotic boundary is located at $z \ra \infty$.
We can do a coordinate transformation: $z=u\sin{\m},~ w=-u\cos{\m}$ resulting in the following metric:
\begin{equation}
    ds^2=\frac{l^2_3}{\sin^2 \m} \left( -\frac{dt^2+du^2}{u^2} +d\m^2\right),
    \label{eq: appendix poincareAdS3}
\end{equation}
where $0<\m<\pi$, and $u>0$.
We can then go to the global patch for each AdS$_2$ slice by doing the following coordinate transformations:
\begin{eqnarray}
    \eta=\ln(u) ,\ \ \ 
    \tau=\left( \frac{2e^{-\eta}}{e^{\eta}+e^{-\eta}} \right)t.
\end{eqnarray}
The metric becomes
\begin{equation}
    ds^2=\frac{1}{\sin^2 \m} \left( -\cosh^2 \left(\frac{\eta}{l_3}\right) d\tau^2 +d\eta^2 +l^2_3d\m^2  \right),
\end{equation}
where $-\infty <\eta <\infty$.
Following a further coordinate transformation below:
\begin{equation}
    y=l_3 \log{\left(\cot{\frac{\m}{2}} \right)},
    \label{eq: appendix y trans}
\end{equation}
we find the metric describing AdS$_2$ global patch slicing of AdS$_3$:
\begin{equation}
    ds^2=dy^2+\cosh^2 \left( \frac{y}{l_3} \right) \left( -\cosh^2{\eta} d\tau^2 +d\eta^2 \right),
\end{equation}
where $-\infty<y<\infty$ and $-\infty<\eta<\infty$. By applying (\ref{eq: appendix y trans}) to the metric (\ref{eq: appendix poincareAdS3}), it is straightforward to transform to \Poincare patch on AdS$_2$ slicing of AdS$_3$:
\begin{equation}
    ds^2= dy^2+l^2_3 \cosh^2 \left( \frac{y}{l_3} \right) \left( -\frac{dt^2+dx^2}{x^2} \right),
\end{equation}
by applying the following coordinate transformations:
\begin{equation}
    t=e^{r_h \tau} \frac{l_3 r}{\sqrt{r^2-r_h^2}},\ \ \ x=e^{r_h \tau} \frac{l_3 r_h}{\sqrt{r^2-r_h^2}},
\end{equation}
we eventually get the AdS$_2$ black hole patch slicing of AdS$_3$:
\begin{equation}
    ds^2= dy^2+\cosh^2 \left( \frac{y}{l_3} \right)\left( -\frac{r^2-r_h^2}{l_3^2}d\t^2+\frac{l_3^2}{r^2-r_h^2}dr^2 \right).
\end{equation}
This metric describes the black string metric in AdS$_3$.

\bibliographystyle{JHEP}

\providecommand{\href}[2]{#2}\begingroup\raggedright\endgroup

\end{document}